\documentclass[a4paper,12pt]{article}
\usepackage{jheppub}
\usepackage[dvipsnames]{xcolor}

\usepackage{pgfplots}
\usepackage[T1]{fontenc}
\usepackage[]{slashed}
\usepackage[]{bm}     
\usepackage{physics}
\usepackage{lipsum}
\usepackage{dsfont}
\usepackage{soul}

\usepackage{graphicx}
\usepackage{epstopdf}
\usepackage{bbm,amsmath,graphicx,amssymb,amsfonts,amsthm}

\usepackage[force]{feynmp-auto}

\definecolor{bubbles}{rgb}{0.91, 1.0, 1.0}
\definecolor{aquamarine}{rgb}{0.5, 1.0, 0.83}
\definecolor{bubblegum}{rgb}{0.99, 0.76, 0.8}
\definecolor{bluebell}{rgb}{0.64, 0.64, 0.82}
\definecolor{dollarbill}{rgb}{0.72, 0.93, 0.6}

\def\sq[#1,#2]{\left[#1\,#2\right]}
\def\an[#1,#2]{\left\langle#1\,#2\right\rangle}

\def\an[#1,#2]{\left\langle#1\,#2\right\rangle}
\def\aq[#1,#2,#3]{\left\langle#1|#2|#3\right]}
\def\qa[#1,#2,#3]{\left[#1|#2|#3\right\rangle}
\def\sq[#1,#2]{\left[#1\,#2\right]}
\def\spa#1.#2{\left\langle#1\,#2\right\rangle}
\def\spab[#1,#2,#3]{\left\langle#1|#2|#3\right]}
\def\spba[#1,#2,#3]{\left[#1|#2|#3\right\rangle}
\def\spb#1.#2{\left[#1\,#2\right]}

\def\sss(#1,#2){s_{{#1}{#2}}}
\def\ttt(#1,#2,#3){s_{{#1}{#2}{#3}}}
\def\uuu(#1,#2,#3,#4){s_{{#1}{#2}{#3}{#4}}}
\def\www(#1,#2,#3,#4,#5){s_{{#1}{#2}{#3}{#4}{#5}}}
\def\eee(#1,#2){(\zeta_{#1}\cdot\zeta_{#2})}
\def\eeek(#1,#2){(\zeta_{#1}\cdot {#2})}
\def\deltabar{{\hat\delta}}

\def\Ttrma(#1,#2,#3,#4){{\rm tr}_{-}[\slash \!\!\!\;\!\! #1\slash  \!\!\!\;\!\! #2 \slash  \!\!\!\;\!\!#3\slash  \!\!\!\;\!\!#4]}
\def\Ttrmb(#1,#2,#3,#4,#5,#6){{\rm tr}_{-}[\slash \!\!\!\;\!\! #1\slash  \!\!\!\;\!\! #2 \slash  \!\!\!\;\!\!#3\slash  \!\!\!\;\!\!#4\slash  \!\!\!\;\!\!#5\slash  \!\!\!\;\!\!#6]}
\def\Ttrmc(#1,#2,#3,#4,#5,#6,#7,#8){{\rm tr}_{-}[\slash \!\!\!\;\!\! #1\slash  \!\!\!\;\!\! #2 \slash  \!\!\!\;\!\!#3\slash  \!\!\!\;\!\!#4\slash  
\!\!\!\;\!\!#5\slash  \!\!\!\;\!\!#6\slash  \!\!\!\;\!\!#7\slash  \!\!\!\;\!\!#8]}
\def\Dp(#1,#2){(#1\cdot #2)}

\def\triangleboxleft{\scalebox{.9}{$\triangleleft$}\kern-.1em\Box}
\def\triangleboxright{\Box\kern-.1em\scalebox{.9}{$\triangleright$}}
\def\dBox{\Box\kern-.1em\Box}
\def\dNPBoxs{\scalebox{.9}{$\bowtie$}\kern-.1em\Box}
\def\dNPBoxu{\Box\kern-.1em\scalebox{.9}{$\bowtie$}}

\def\beq{\begin{equation}}
\def\eeq{\end{equation}}
\def\bes{\begin{split}}
\def\ees{\end{split}}
\def\beqa{\begin{eqnarray}}
\def\eeqa{\end{eqnarray}}

\def\eeqa{\end{eqnarray}}

\pgfdeclarelayer{bg}    % declare background layer
\pgfsetlayers{bg,main}  % set the order of the layers (main is the standard layer)

\definecolor{Mathematica}{HTML}{ed192d}

\usepackage{float}

      \usepackage[T1]{fontenc} % if needed

\title{\boldmath Post-Minkowskian Radial Action from Soft Limits and Velocity Cuts}

\author[a]{N. Emil J. Bjerrum-Bohr,}
\author[a]{Ludovic Plant\'e,}
\author[b,c]{and Pierre Vanhove}

\preprint{IPhT-t21/072}

\affiliation[a]{Niels Bohr International Academy, Niels Bohr Institute, University of Copenhagen, Blegdamsvej 17, DK-2100 Copenhagen, Denmark}
\affiliation[b]{Institut de Physique Theorique, Universit\'e Paris-Saclay,
CEA, CNRS, F-91191 Gif-sur-Yvette Cedex, France}
\affiliation[c]{National Research University Higher School of
  Economics, Russian Federation}
\abstract{We consider gravitational massive scalar-scalar scattering
 from unitarity and demonstrate how intermediate soft graviton
 behavior and the concept of extracting classical physics from
 localization of integrands on velocity cuts devise an efficient
 extraction scheme for computing the classical post-Minkowskian
 radial action perturbatively. We demonstrate the computational
 efficiency by deriving the scattering amplitudes in the probe regime
 to the fifth post-Minkowskian order in arbitrary dimensions.  }
\begin{document} 
\maketitle
\flushbottom
\section{Introduction}
In an age where gravitational waves in the Universe can be witnessed from black hole and neutron star mergers~\cite{LIGOScientific:2016aoc,LIGOScientific:2017vwq}, an exciting particle physics theme is computation of relativistic classical interactions from gravitational quantum scattering ~\cite{Damour:2016gwp,Damour:2017zjx}. Here modern amplitudes techniques are handy~\cite{Neill:2013wsa,Bjerrum-Bohr:2013bxa,Bjerrum-Bohr:2018xdl,Cheung:2018wkq}, and adaptable for the provision of precision predictions in gravity at relativistic velocities~\cite{Bern:2019nnu,Antonelli:2019ytb,KoemansCollado:2019ggb,Cristofoli:2019neg,Bern:2019crd, Kalin:2019rwq,Bjerrum-Bohr:2019kec,Bern:2020gjj,Cristofoli:2020uzm, Cheung:2020gyp,Parra-Martinez:2020dzs,DiVecchia:2020ymx,Mougiakakos:2020laz,Bern:2021dqo,Bjerrum-Bohr:2021vuf,Bjerrum-Bohr:2021din,Brandhuber:2021eyq,Damgaard:2021ipf} (for some applications with spin see for instance~\cite{Vaidya:2014kza,Maybee:2019jus,Chung:2019duq,Chung:2020rrz}). \\[5pt]
At heart of exploring classical general relativity from quantum
scattering amplitudes~\cite{Iwasaki:1971vb} is the correspondence
principle of quantum mechanics~\cite{Bohr:1920}. It stipulates the
emergence of classical observables when quantum numbers are large, and
naturally, an ideal application is gravitational quantum scattering
amplitudes for superheavy black hole point particles~\cite{Feynman:1963ax,DeWitt:1967yk,DeWitt:1967ub,DeWitt:1967uc},
in context of Weinberg's~\cite{Hawking:1979ig} widely celebrated idea of general
relativity as a low-energy effective field
theory~\cite{Donoghue:1993eb,Donoghue:1994dn,BjerrumBohr:2002kt,BjerrumBohr:2002ks,BjerrumBohr:2004mz}. \\[5pt]
Current applications revolve around obtaining low-energy quantum
$S$-matrix elements in an asymptotic Minkowskian flat background
arranged in powers of Newton's constant ${\cal O}(G_N)$, ${\cal
 O}(G_N^2),\ldots$. Thus deriving $L+1$ post-Minkowskian order classical physics require $L$-loop scattering amplitudes~\cite{Holstein:2004dn,Bjerrum-Bohr:2013bxa,Bjerrum-Bohr:2018xdl,Kosower:2018adc}. \\[5pt]
In such computations (and contrasting most other precision physics
loop amplitude computations), only long-distance (non-analytic)
components with varying orders of Planck's constant $\hbar$ have to be
computed
\cite{Donoghue:1993eb,Donoghue:1994dn,Bjerrum-Bohr:2013bxa,Bjerrum-Bohr:2018xdl}. To
compute a classical Hamiltonian from a scattering amplitude, we can use either Born subtractions in the context of the Lippmann-Schwinger equation~\cite{Cristofoli:2019neg, Kalin:2019rwq,Bjerrum-Bohr:2019kec}, or equivalently an effective field theory matching procedure~\cite{Cheung:2018wkq,Bern:2019nnu,Bern:2019crd,Cheung:2020gyp,Bern:2021dqo}. 
At low orders in perturbation, this computational scheme is very
efficient, but each new loop order poses a challenge in part because of the new integrals involved, but also since the number of amplitude pieces that have to be identified, discarded, subtracted, and kept growing factorially with the perturbative order considered. \\[5pt]
The purpose of this paper is to devise a refined computational
technology that allows focusing on exactly those amplitude integrand
components that integrate to the classical radial action. This work
generalises to all loop orders  the understanding of how classical physics appear in
such computations at one- and two-loop
orders~\cite{Bjerrum-Bohr:2021vuf,Bjerrum-Bohr:2021din}. 
The method uses the {\it velocity cuts} formalism introduced in
refs.~\cite{Bjerrum-Bohr:2021vuf,Bjerrum-Bohr:2021din}.
It was shown that reorganizing some combination of propagators in the integrand (see for example Eqn.~(3.4) and~(3.6) in ref.~\cite{Bjerrum-Bohr:2021din})
\begin{multline} \label{propid}
\left(\frac{1}{({p}_a \cdot
 \ell_a\!+\!i\varepsilon)({p}_a \cdot \ell_b\!-\!i\varepsilon)}\!-\!\frac{1}{({p}_a
 \cdot \ell_b\!+\!i\varepsilon)({p}_a \cdot
 \ell_a\!-\!i\varepsilon)}\right)\times\cr
 \left(\frac{1}{({p}_b \cdot \ell_a\!-\!i\varepsilon)({p}_b \cdot \ell_c\!+\!i\varepsilon)}\!-\!\frac{1}{({p}_b \cdot \ell_c\!-\!i\varepsilon)({p}_b \cdot \ell_a\!+\!i\varepsilon)}\right),
\end{multline}
in terms of delta functions
\begin{equation}
\left(\frac{\delta({p}_a \cdot \ell_a)}{{p}_a \cdot \ell_b+i\varepsilon}-\frac{\delta({p}_a \cdot \ell_b)}{{p}_b \cdot \ell_a+i\varepsilon}\right)\times\left(\frac{\delta({p}_b \cdot \ell_c)}{{p}_b \cdot \ell_a+i\varepsilon}-\frac{\delta({p}_b \cdot \ell_a)}{{p}_b \cdot \ell_c+i\varepsilon}\right),
\end{equation}
before integration, lead to considerable simplifications of the classical part of one- and two-loop amplitude computations.
We use a multi-soft graviton expansion to organize the
$\hbar\to0$ expansion of the integrand. This allows us to collect 
terms in the integrand according to unitarity relations in the
exponential representation of the $S$-matrix
of~\cite{Damgaard:2021ipf}. 
This organization of the multi-loop integrand combined with
the velocity cut formalism provides a direct identification of the
classical radial action $N_L$ at each loop order.\\[5pt]
As we will see this saves computational resources, and we will
demonstrate efficiency by direct calculation of the probe limit of the
radial action until the fifth post-Minkowskian order.
We will verify calculations by deriving the scattering angle in the
probe limit in arbitrary dimensions, and compare to known results derived from the Schwarzschild-Tangherlini metric.\\[5pt] 
 The paper is organized as follows. 
First, we review needed background knowledge for computations and the  
gravitational coupling of scalar massive fields in the context of the
Einstein-Hilbert Lagrangian. This is followed by the section~\ref{sec:CHY} where 
we give compact expressions for the scalar-graviton tree-level amplitudes, that will
be used in the loop amplitude computations. In section~\ref{sec:unitaritycuts} we
give a new organization of the tree-level 
scalar-multi-graviton amplitude, and in section~\ref{sec:multisoft} we show how it is possible to
manifest the multi-soft behaviour in a way that allows us to devise a
direct extraction scheme for classical physics from loop-integrands
using velocity cuts. In
section~\ref{sec:probeamp} we evaluate the multi-loop two-body
scattering amplitudes in the probe limit up to the fifth
post-Minkowskian order. Section~\ref{sec:conclusion} contains our
conclusion. Details of the numerators factors and tree-level
amplitudes are given in Appendix~\ref{sec:YMtrees}.
%%%%%%%%%%%%%%%%%%%%%%%%%%%%%%%%%%%%%%%%%%%%%%%%%%%%%%%%%%%%%%%%% 
\section{Classical physics from quantum amplitudes}
\label{sec:3PMlevel}
%%%%%%%%%%%%%%%%%%%%%%%%%%%%%%%%%%%%%%%%%%%%%%%%%%%%%%%%%%%%%%%%
We will focus on the minimal gravitational coupling of scalar fields $(\phi_1,\phi_2)\to(\phi_1,\phi_2)$ with masses $m_1$ and $m_2$ from the effective field theory Einstein-Hilbert Lagrangian,
\begin{equation}
{\cal L} = \int d^4 x \sqrt{-\det(g)} \Bigg[\frac{R}{16 \pi G_N} + \frac12 g^{\mu\nu} ( \partial_\mu \phi_1\partial_\nu \phi_1 + \partial_\mu \phi_2\partial_\nu \phi_2)- m_1^2 \phi_1^2 - m_2^2 \phi_2^2\Bigg]+{\cal L}_{\rm EFT}\,,
\end{equation} 
here ${\cal L}_{\rm EFT}$ denote effective field theory operators necessary to define a well-behaved low-energy quantum gravity theory at any perturbative order, $G_N$ is Newton's constant, $R$ is the Ricci 
scalar, and the expansion of the metric is defined from $g_{\mu\nu}(x)\equiv \eta_{\mu\nu} + \sqrt{32 \pi G_N}h_{\mu\nu}(x)$ where $\eta_{\mu\nu}$ is the mostly minus Minkowskian metric.\\[5pt]
To extract classical physics from quantum amplitudes, we consider
scalar four-point $L$-loop scattering processes. We derive such
amplitudes from $(L+1)$-graviton generalised unitarity
cuts. Computations are organized in a perturbative expansion in
Newton's constant $G_N$.
\begin{equation}
{\mathcal M}(p_1,p_2,p_1',p_2')=
\sum_{L=0}^{ \infty} \mathcal{M}_{ L}(p_1,p_2,p_1',p_2') =
\begin{gathered}
  \begin{fmffile}{other9}
 \begin{fmfgraph*}(100,100)
\fmfleftn{i}{2}
\fmfrightn{o}{2}
\fmfv{decor.shape=oval, decor.filled=shaded, decor.size=(.5w)}{v1}
\fmfv{decor.shape=circle,decor.filled=gray50,decor.size=(.5w)}{v1}
\fmf{fermion,label=$p_1$,label.side=left}{i1,v1}
\fmf{fermion,label=$p_1'$,label.side=left}{v1,i2}
\fmf{fermion,label=$p_2'$,label.side=right}{v1,o2}
\fmf{fermion,label=$p_2$,label.side=right}{o1,v1}
\end{fmfgraph*}
\end{fmffile}
\end{gathered}
\,,
\end{equation}
where $ \mathcal{M}_{ L}$ denotes the $L$-loop scalar four-point
scattering process of order $G_N^{L+1}$. We will suppress the
Newton constant and only reinstate it in section~\ref{sec:probeamp}
where we compute the probe radial action.
We have defined $p_1$ and $p_2$ and ${p_1'}$ and ${p_2'}$ as on-shell
incoming and outgoing momenta respectively, and
$p_1^2 \equiv {p_1'}^2 \equiv m_1^2$, $p_2^2\equiv {p_2'}^2\equiv
m_2^2$.
The center-of-mass energy is $E_{CM}^2=s\equiv(p_1+p_2)^2
\equiv({p_1'}+{p_2'})^2$, and we introduce the relativistic factor $\sigma\equiv
\frac{p_1 \cdot p_2}{m_1 m_2} $. The
transfer momentum is $q=p_1-p_1'=p_2'-p_2$. We remark that $q\cdot
p_1=-q\cdot p_1'={q^2\over2}$.
%
%%%%%%%%%%%%%%%%%%%%%%%%%%%%%%%%%%%%%%%%%%%%%%%%%%%%%%%%%%%%%
%

We will consider $L$-loop integrands from generalised $(L+1)$ graviton unitarity 
cuts\footnote{From now we
will remove the polarisation label, and indicate it only when
evaluating the cut in section~\ref{sec:probeamp}.}  (We refer to~\cite{Bjerrum-Bohr:2021vuf} for the $\hbar$ factors in the
loop amplitudes.)
\begin{multline}\label{e:MLcut}
 i\mathcal M_{L}^{\textrm{cut}} (\sigma,q^2)=\hbar^{3L+1} \int (2\pi)^D\delta(q+\ell_2+\cdots+\ell_{L+2})\prod_{i=2}^{L+2} {i
 \over\ell_i^2}\prod_{i=2}^{L+2} { d^D\ell_i\over (2\hbar\pi)^D} 
 \cr{1\over (L+1)!}\sum_{h_i=\pm2}
 M^{\rm tree}_{\rm Left}(p_1,\ell_2^{h_2},\ldots,\ell_{L+2}^{h_{L+2}},-p_1')
 M^{\rm tree}_{\rm Right}(p_2,-\ell_2^{h_2},\ldots,-\ell_{L+2}^{h_{L+2}},-p_2')^\dagger,
\end{multline}
where $M^{\rm tree}_{\rm Left}(p_1,\ell_2,\ldots,\ell_{L+2},-p_1')$ and
$M^{\rm tree}_{\rm Right}(p_2,-\ell_2,\ldots,-\ell_{L+2},-p_2')$ are tree-level
amplitudes of 
multi-graviton emission from a massive scalar. We take the
convention that all graviton lines are incoming in the left tree-level
factor and out-going in the right tree-level factor, with the momentum
conservation 
\begin{equation}
 q=p_1-p_1'=-\sum_{i=2}^{L+2}\ell_i. 
\end{equation}

\begin{figure}\centering\begin{fmffile}{bowtie}
\begin{fmfgraph*}(100,100)
\fmfstraight
\fmfleftn{i}{2}
\fmfrightn{o}{2}
\fmf{fermion,label=$p_1$,label.side=left}{i1,v1}
\fmf{fermion,label=$p_1'$,label.side=left}{v2,i2}
\fmf{fermion,label=$p_2'$,label.side=right}{v3,o2}
\fmf{fermion,label=$p_2$,label.side=right}{o1,v4}
\fmf{fermion,tension=0.01}{v1,v2}
\fmf{fermion,tension=0.01}{v4,v3}
\fmf{dbl_wiggly}{v1,v5,v3}
\fmf{dbl_wiggly}{v2,v5,v4}
\fmfblob{6thick}{v5}
\fmf{phantom,tension=0}{v1,v2}
\fmf{phantom,tension=0}{v4,v3}
\end{fmfgraph*}
\end{fmffile}
\caption{The bowtie diagram that arises at two loops.}
\label{fig:bowtie}
\end{figure}
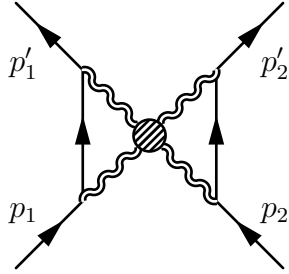

The multi-graviton cut is not enough for reconstructing the full 
classical $L$-loop amplitude. We need to add terms that are not contained in the cut
in eq.~\eqref{e:MLcut}. A first type of contributions are multi-graviton
cuts factorising the amplitude into a product of two scalar-graviton
amplitudes  times graviton amplitudes
\begin{multline}
    M^{\rm tree}_{\rm
      Left}(p_1,\ell_2,\ldots,\ell_{n},-p_1')\times M^{\rm grav.}(-\ell_2,\ldots,-\ell_{n},\ell_{n+1},\ldots,\ell_{m})\cr\times
 M^{\rm tree}_{\rm Right}(p_2,-\ell_{n+1},\ldots,\ell_{m},-p_2')^\dagger,
\end{multline}
where $M^{\rm grav.}$ is a pure gravity amplitude. Such a contribution
arises at two-loop order from the bow-tie graph in fig.~\ref{fig:bowtie}
for which $M^{\rm grav.}$ is
a four-graviton tree amplitude. Since the construction of the
integrands presented in this paper is an organisation the
scalar-graviton 
tree-level amplitudes, there is no obstacle from including such contributions.

A second type of contributions are
the self-energy ones (see section~4
of~\cite{Bjerrum-Bohr:2021din}
and~\cite{Parra-Martinez:2020dzs,Herrmann:2021lqe}). After  cutting all
the graviton lines these amplitudes factorise two tree-level
scalar-graviton amplitudes times possibly multi-graviton
amplitudes. Again the manipulations of the scalar-graviton tree
amplitude presented in this paper can be applied to this case, but their analysis is beyond the scope of the present paper.

%----------------------------------------------------------------------
\section{Compact Massive-Scalar-Graviton Tree Amplitudes}\label{sec:CHY}
In this section, we give the tree-level multi-graviton
emission from a scalar line. We use the scattering equations formalism
\cite{Cachazo:2013hca,Cachazo:2013iea,Cachazo:2013gna,Cachazo:2014nsa}
which derives amplitudes for a large class of field theories through moduli
integrations over string-theory-like integrands in very compact ways. 
Systematic algebraic construction of numerators for gluon in the
scattering equation framework was pioneered by Fu, Du, Huang, and Feng
in \cite{Fu:2017uzt} and in ref. \cite{Teng:2017tbo,Bjerrum-Bohr:2020syg} developed into
diagrammatic methods.

\noindent
Provided the scattering equation prescription for the colour-ordered
multi-gluon amplitude in~\cite{Cachazo:2013iea,He:2017spx} and 
following the construction in
\cite{Bjerrum-Bohr:2020syg} for 
tree-level amplitudes, the Yang-Mills multi-gluon amplitude 
(with Yang-Mills coupling at unity) has the form

\begin{multline}
{A}_{n-2}(1,\{2,\dots,n-1\},n)= \int {\prod_{i=1}^n dz_i \over
 \textrm{vol}(SL(2,\mathbb C))} \prod_{i=1}^n
\delta'\left(\sum_{j=1\atop j\neq i}^n{k_i\cdot k_j\over z_{ij}}\right)
\frac{1}{z_{1 2}\cdots z_{{n-1}\, {n}} }\cr
\times\sum_{\beta\in \mathfrak S_
 {n-2}}{N_{n-2}(1,\beta(2,\dots,n-1),n)\over z_{1 \beta(2)}z_{\beta(2)\beta(3)}\cdots z_{\beta(n-1)\, {n}} }\,,
\end{multline}
where we have made use of the notation $z_{ij}=z_i-z_j$, and $N_{n-2}(1,\beta(2,\dots,n-1),n)$ are numerator expressions containing $i$th gluon polarisation vectors $\zeta_i$ and momenta $k_i$. In contrast to the numerators developed in \cite{Fu:2017uzt} we will as suggested in \cite{Bjerrum-Bohr:2020syg} average over reference orderings since it provides a computational advantage. Integration over the scattering equations can be done numerically but for arriving at analytic expressions 
one can explore the one-to-one link between integration measures 
and integrands in
scattering equations and traditional string
theory~\cite{Bjerrum-Bohr:2014qwa} and use it to formulate
scattering equation integration
rules~\cite{Baadsgaard:2015voa,Baadsgaard:2015ifa,Bjerrum-Bohr:2016juj,Bjerrum-Bohr:2016axv}.\\[5pt] 
Following this analytic procedure, we arrive at the following expression for the amplitude 
 \begin{equation}\begin{split}
& A_{n-2}(1,\beta(2,\dots,n-1),n)=\ \ \ \ \ \ \ \\&\ \sum_{\gamma\in\mathfrak S_{n-2}}\!\!\!\!
m^{\rm tree}(1,\beta(1,\dots,n),n| 1,\gamma(2,\dots,n-1),n) N_{n-2}(1,\gamma(2,\dots,n-1),n)\,,  
\end{split} \end{equation}
 where $m^{\rm tree}(1,\beta(2,\dots,n-2),n,n-1|
 1,\gamma(2,\dots,n-2),n-1,n)=\mathcal S^{-1}(\beta|\gamma)|_{p_1}$ can be interpreted as the inverse momentum kernel as demonstrated in~\cite{Cachazo:2013iea}. Plugging this in the momentum kernel expression for gravity amplitudes~\cite{Bjerrum-Bohr:2010diw,Bjerrum-Bohr:2010kyi,Bjerrum-Bohr:2010mtb,Bjerrum-Bohr:2010pnr} immediately yields
 \begin{equation}\begin{split}
 & M^{\rm tree}_{n-2}(1,2,\dots,n\!-\!1,n)=\\ 
 &\!\! (-1)^n\!\!\!\!\!\!\! \sum_{\beta,\gamma\in\mathfrak
  S_{n-3}}\!\!\!\!\!\!\!A_{n-2}(1,\beta(2,\dots,n\!-\!2),n\!-\!1,n)
\mathcal S(\beta|\gamma)|_{p_1} A_{n-2}(p_1,n\!-\!1,\gamma(2,\dots,n\!-\!2),n)\,,  
 \end{split} \end{equation}
leading to the following compact prescription for computing tree-level multi-graviton amplitudes
\begin{equation}\label{e:MtreeKLTCHY}
{M}^{\rm tree}_{n-2}(1,2,\ldots,n) = 
i\,\sum_{\beta\in \mathfrak S_ {n-2}} N_{n-2}(1,\beta(2,\cdots,n-1),n) {
 A}_{n-2}(1,\beta(2,\dots,n-1),n) \,,
\end{equation}
where ${
 A}_{n-2}(1,\gamma(2,\dots,n-1),n)$ are the colour-ordered multi-gluon
tree-level amplitudes.
As explained in \cite{Cachazo:2013iea,Bjerrum-Bohr:2020syg} we can
derive numerators with gluons states replaced by massive scalar states
$k_1 \to p$, $k_n \to -p'$ by dimensional reduction with $p^2=(p')^2=m^2$. This has the
effect of replacing the multi-gluon numerator factors
$N_{n-2}(1,\beta(2,\dots,n-1),n)$ by $N_{n-2}(p,\beta(2,\dots,n-1),-p')$ so
that the amplitude for multi-gluon emission from a massive scalar reads
\begin{multline}
  A_{n-2}(p,\beta(2,\dots,n-1),-p')=\cr\sum_{\gamma\in\mathfrak S_{n-2}}\!\!\!\!
m^{\rm tree}(p,\beta(2,\dots,n-1),-p'| p,\gamma(2,\dots,n-1),-p') N_{n-2}(p,\gamma(2,\dots,n-1),-p')\,.
 \end{multline}
Thus the amplitude for emission of gravitons from a massive scalar can be written
\begin{multline}
{M}^{\rm tree}_{n-2}(p,2,\ldots,n-1,-p') = \cr
i\,\sum_{\beta \in \mathfrak S_ {n}}  N_{n-2}(p,\beta(2,\dots,n-1),-p') {
 A}_{n-2}(p,\beta(2,\dots,n-1),-p') \,.
\end{multline}
The numerator factors for the scalar multi-gluon amplitudes we will use here are
constructed using the method of ref.~\cite{Bjerrum-Bohr:2020syg}. The
expressions for the numerator factors and Yang-Mills tree-level
amplitudes are collected in appendix~\ref{sec:YMtrees}.
A consequence of the colour-kinematics representation provided is that the numerator factors for gravity amplitudes are perfect squares of linear combinations of Yang-Mills numerators. We summarise the required amplitudes below.
\begin{itemize}
\item At three-point order we have 
\begin{equation} \label{e:M3pt}
M^{\rm tree}_{1}(p,\ell_2,-p')=i\,N_1(p,\ell_2, -p')A_1(p,\ell_2, -p')=i\,N_1(p,\ell_2, -p')^2 ,
\end{equation}
where the numerator is given in appendix~\ref{sec:3pt}.
\item At four-points order we have
\begin{equation} \label{e:M4pt}\begin{split}
M^{\rm tree}_{2}(p,\ell_2,\ell_3,-p')
= i\,N_{2}(p,2,3,-p') {A}_{2}(p,2,3,-p')+ {\rm perm.} \{2,3\}\\ 
=\frac{{i N_2}(p,2,3,-p')^2}{(\ell_2+p)^2-m^2+i\varepsilon}+ \frac{{i N_2}(p,3,2,-p')^2}{(\ell_3+p)^2-m^2+i\varepsilon}+\frac{i({N_2}^{[2,3]})^2}{(\ell_2+\ell_3)^2+i\varepsilon}\,,
\end{split}\end{equation}
with ${N_2}^{[2,3]}\equiv {N_2}(p,2,3,-p')-{N_2}(p,3,2,-p')$. The numerator is given in appendix~\ref{sec:4pt}.
\item At five-point order we have 
\begin{multline}\label{M5pt}
M^{\rm tree}_{3}(p,\ell_2,\ell_3,\ell_4,-p') = i\, N_{3}(p,2,3,4,-p') {A}_{3}(p,2,3,4,-p')+ {\rm perm.} \{2,3,4\}\cr 
= \frac{i\, ({N_3}^{2,3,4})^2}{((p+\ell_2)^2-m^2+i\varepsilon)((p+\ell_2+\ell_3)^2-m^2+i\varepsilon)
 } \cr
 +
 \frac{ i( {N_3}^{[2,3],4})^2}{2((p+\ell_2+\ell_3)^2-m^2+i\varepsilon)
   ((\ell_2+\ell_3)^2+i\varepsilon)} \cr
 +
 \frac{ i({N_3}^{2,[3,4]})^2}{2((p+\ell_2)^2-m^2+i\varepsilon)
  ((\ell_3+\ell_4)^2+i\varepsilon) }\cr
 +\frac{ i({N_3}^{[2,[3,4]]})^2}{4((\ell_2+\ell_4)^2+i\varepsilon) ((\ell_2+\ell_3+\ell_4)^2+i\varepsilon)} 
 \cr+\frac{ i({N_3}^{[[2,3],4]})^2}{4((\ell_2+\ell_3)^2+i\varepsilon) ((\ell_2+\ell_3+\ell_4)^2+i\varepsilon)
 } + {\rm perm.} \{2,3,4\}\,,
\end{multline}
where the numerator is given in appendix~\ref{sec:5pt}.
\item For six-point we have
\begin{multline} 
{M}^{\rm tree}_{4}(p,\ell_2,\ell_3,\ell_4,\ell_5,-p')
=i\, N_{4}(p,2,3,4,5,-p') {A}_{4}(p,2,3,4,5,-p')\cr+ {\rm perm.}
\{2,3,4,5\},
\end{multline}
which gives 
\begin{multline} \label{M6pt}
{M}^{\rm tree}_{4}(p,\ell_2,\ell_3,\ell_4,\ell_5,-p')
=\cr
\frac{i\,({N_4}^{2,3,4,5})^2}{\sss(2,p) \ttt(2,3,p) \uuu(2,3,4,p)}
+\frac{{}i({N_4}^{2,3,[4,5]})^2}{2\sss(4,5) \sss(2,p)
 \ttt(2,3,p)}
+\frac{{}i({N_4}^{2,[3,4],5})^2}{2\sss(3,4) \sss(2,p)
 \uuu(2,3,4,p)}
+\frac{{}i({N_4}^{[2,3],4,5})^2}{2\sss(2,3) \ttt(2,3,p)
 \uuu(2,3,4,p)}\cr 
+\frac{{}i({N_4}^{[2,3],[4,5]})^2}{4\sss(2,3) \sss(4,5)
 \ttt(2,3,p)}
+\frac{{}i({N_4}^{[[2,3],4],5})^2}{4\sss(2,3) \ttt(2,3,4)
 \uuu(2,3,4,p)}
+\frac{{}i({N_4}^{[2,[3,4]],5})^2}{4\sss(3,4) \ttt(2,3,4)
 \uuu(2,3,4,p)}\cr
+\frac{i({N_4}^{2,[[3,4],5]})^2}{4\sss(3,4) \ttt(3,4,5)
 \sss(2,p)}
+\frac{i({N_4}^{2,[3,[4,5]]})^2}{4\sss(4,5) \ttt(3,4,5)
 \sss(2,p)}\cr 
+\frac{i({N_4}^{[[2,3],[4,5]]})^2}{8\sss(2,3) \sss(4,5)
 \uuu(2,3,4,5)}
+ \frac{i({N_4}^{[[[2,3],4],5]})^2}{8\sss(2,3)
 \ttt(2,3,4)
 \uuu(2,3,4,5)}
+ \frac{i({N_4}^{[[2,[3,4]],5]})^2}{8\sss(3,4)
 \ttt(2,3,4)
 \uuu(2,3,4,5)}\cr
+ \frac{i({N_4}^{[2,[[3,4],5]]})^2}{8\sss(3,4)
 \ttt(3,4,5)
 \uuu(2,3,4,5)}
+ \frac{i({N_4}^{[2,[3,[4,5]]]})^2}{8\sss(4,5)
 \ttt(3,4,5) \uuu(2,3,4,5)}+{\rm perm.} \{2,3,4,5\},
\end{multline}
%.
where we have used the short-hand notations
$s_{i_1,\dots ,i_r, p}=(p+\sum_{j=1}^r\ell_{i_r})^2-m^2+i\varepsilon$,
$s_{i_1,\dots,i_r}=(\sum_{j=1}^r \ell_{i_j})^2+i\varepsilon$. The Yang-Mills amplitude is given in appendix~\ref{sec:6pt}.
\end{itemize}
With similar expressions for the seven-point amplitude, where the Yang-Mills amplitude is given in the appendix~\ref{sec:7pt}.
%
%----------------------------------------------------------------------
\section{Unitarity and multi-graviton emission}\label{sec:unitaritycuts}
We consider the amplitude in~\eqref{e:MLcut} with   $L+1$ 
graviton generalised unitarity cuts.
We will make use of the notation 
\begin{equation}\label{e:MLhat}
 M^{\rm tree}_{\rm Left}(p_1,\ell_2,\dots,\hat\ell_i,\ldots,\ell_{L+1},{\ell}_{L+2},-p_1'),
\end{equation}
where we indicate the leg on which we have used momentum conservation
by a hat 
\begin{equation}
 \ell_i=-q-\sum_{2\leq j\leq L+2\atop j\neq i}\ell_j.
\end{equation}
The tree-level amplitudes in~\eqref{e:MLhat} have two kinds of massive
propagators. The propagators containing a `hatted' leg read 
\begin{multline}
\frac{1}{(p_1-\ell_{i_2}-\cdots-\ell_{i_j}-q)^2-m^2+i \varepsilon}\cr=\frac{-1}{2p_1\cdot(\ell_{i_2}+\cdots+\ell_{i_j}) -(\ell_{i_2}+\cdots+\ell_{i_j}+q)^2-i \varepsilon},
\end{multline}
while the propagators which do not involve the `hatted' leg are
\begin{equation}
\frac{1}{(p_1+\ell_{i_2}+\cdots+\ell_{i_j})^2-m^2+i \varepsilon}=\frac{1}{2p_1\cdot(\ell_{i_2}+\cdots+\ell_{i_j}) +(\ell_{i_2}+\cdots+\ell_{i_j})^2+i \varepsilon}\,.
\end{equation}
In the above equations we need to take $1<j\leq L+2$.\\[5pt]
Using the identity
\begin{equation}\label{e:PP}
 \lim_{\varepsilon\to0^+}\left( {1\over \eta-i\varepsilon}-{1\over \eta+i\varepsilon}\right)=
 \lim_{\varepsilon\to0^+} {2i\varepsilon \over \eta^2+\varepsilon^2}=
\deltabar(\eta),
\end{equation}
we rewrite propagators with a `hatted' leg 
\begin{multline}\label{e:propflip}
\frac{1}{(p_1-\ell_{i_2}-\cdots-\ell_{i_j}-q)^2-m^2+i \varepsilon}=
\deltabar\big((p_1-\ell_{i_2}-\cdots-\ell_{i_j}-q)^2-m^2\big)\cr
+\frac{1}{(p_1-\ell_{i_2}-\cdots-\ell_{i_j}-q)^2-m^2-i \varepsilon}\,,
\end{multline}
where we have introduced the notation $\deltabar(x) \equiv-2\pi i
\delta(x)$.

We denote by
$M^{\textrm{tree}(+)}_{L+1}(p_1,\ell_2$ $, \dots,{\ell}_{L+2},-p_1')$ the tree-level amplitudes where all the propagators of the type $2p_1\cdot (\sum_r \ell_r)- (q+\sum_r \ell_r)^2-i
\varepsilon$ are flipped to $2p_1\cdot (\sum_r \ell_r)- (q+\sum_r
\ell_r)^2+i \varepsilon$.
We denote $M^{\textrm{tree}(-)}_{L+1}(p_1,\ell_2,\dots,{\ell}_{L+2},-p_1')$ the tree-level
amplitudes where all the propagators of the type $2p_1\cdot (\sum_r \ell_r)+ (\sum_r \ell_r)^2+i \varepsilon$ are flipped to
$2p_1\cdot (\sum_r \ell_r)+(\sum_r \ell_r)^2-i \varepsilon$.

\medskip

In the next sections, we focus first on two- and three-graviton emission
tree-level amplitudes and the relations between the amplitudes $M^{\rm
  tree}_{L+1}$,
$M^{\textrm{tree}(+)}_{L+1}$ and $M^{\textrm{tree}(-)}_{L+1}$ with $L=1$ and $L=2$ followed by a generalisation to generic multi-graviton emission.
%-------------------------------------------------------------------
\subsection{The four-point case}\label{sec:fourpointtree}
We write the four-point Feynman tree-level amplitude in the form
\begin{equation}
 M^{\rm tree}_2(p_1,\hat\ell_2,\ell_3,-p_1')= \frac{n_{p_1+\hat\ell_2}}{(p_1+\hat\ell_2)^2-m_1^2+i \varepsilon}+ \frac{n_{p_1+\ell_3}}{(p_1+\ell_3)^2-m_1^2+i \varepsilon}+ \frac{n_{q}}{q^2},
\end{equation}
with generic (off-shell) numerator factors $n_{p_1+\ell_2}$,
$n_{p_1+\ell_3}$ and $n_{q}$ and where momentum conservation is
imposed on leg $\ell_2=-\ell_3-q$. 
Applying the relation~\eqref{e:propflip} on the first propagator only,
the amplitude reads
\begin{equation}
 M^{\rm tree}_2(p_1,\hat\ell_2,\ell_3,-p_1')= 
 \deltabar((p_1+\hat\ell_2)^2-m_1^2) n_{p_1+\hat\ell_2}+ M^{\textrm{tree}(+)}_2(p_1,\hat\ell_2,\ell_3,-p_1'),
\end{equation}
with 
\begin{equation}\label{e:M2plus}
   M^{\textrm{tree}(+)}_2(p_1,\hat\ell_2,\ell_3,-p_1')\equiv -\frac{n_{p_1-q-\ell_3}}{2p_1\cdot (q+\ell_3)+i \varepsilon}+
\frac{n_{p_1+\ell_3}}{2 p_1\cdot \ell_3+i \varepsilon}+ \frac{n_{q}}{q^2}.
\end{equation}
Using the factorisation theorem on the pole
$(p_1+\hat\ell_2)^2-m_1^2=0$
\begin{equation}
  M^{\rm tree}_2(p_1,\hat\ell_2,\ell_3,-p_1')\sim
  M^{\rm tree}_1(p_1,\hat\ell_2,-p_1-\hat\ell_2)  {i\over
    (p_1+\hat\ell_2)^2-m_1^2}  M^{\rm tree}_1(p_1+\hat\ell_2,\ell_3,-p_1'),
\end{equation}
implies that the support of the delta-function the numerator factor
$n_{p_1+\hat\ell_2}$ factorises into the product of two tree-level
amplitudes. The tree amplitude takes the form (with the cut conditions $\hat\ell_2^2=\ell_3^2=0$)
\begin{multline}\label{e:M2plus}
M^{\rm tree}_2(p_1,\hat\ell_2,\ell_3,-p_1')= M^{\textrm{tree}(+)}_2(p_1,\hat\ell_2,\ell_3,-p_1')\cr
+\deltabar((p_1+\hat\ell_2)^2-m_1^2)
 M^{\rm tree}_1(p_1,\hat\ell_2,-p_1-\hat\ell_2)
 M^{\rm tree}_1(p_1+\hat\ell_2,\ell_3,-p_1').
\end{multline}
Similarly, by flipping the other propagator we have
\begin{multline}\label{e:M2minus}
M^{\rm
  tree}_2(p_1,\hat\ell_2,\ell_3,-p_1')=M^{\textrm{tree}(-)}_2(p_1,\hat\ell_2,\ell_3,-p_1')\cr
+\deltabar((p_1+\ell_3)^2-m_1^2)
M^{\rm tree}_1(p_1,\ell_3,-p_1-\ell_3) M_1(p_1+\ell_2,\hat\ell_2,-p_1'),
\end{multline}
with 
\begin{equation}
    M^{\textrm{tree}(-)}_2(p_1,\hat\ell_2,\ell_3,p_1') \equiv
    -\frac{n_{p_1-q-\ell_3}}{2p_1\cdot (q+\ell_3) -i \varepsilon}+
\frac{n_{p_1+\ell_3}}{2 p_1\cdot \ell_3-i \varepsilon}+ \frac{n_{q}}{q^2}.
\end{equation}
\begin{figure}[ht]
 \centering
\includegraphics[width=12cm]{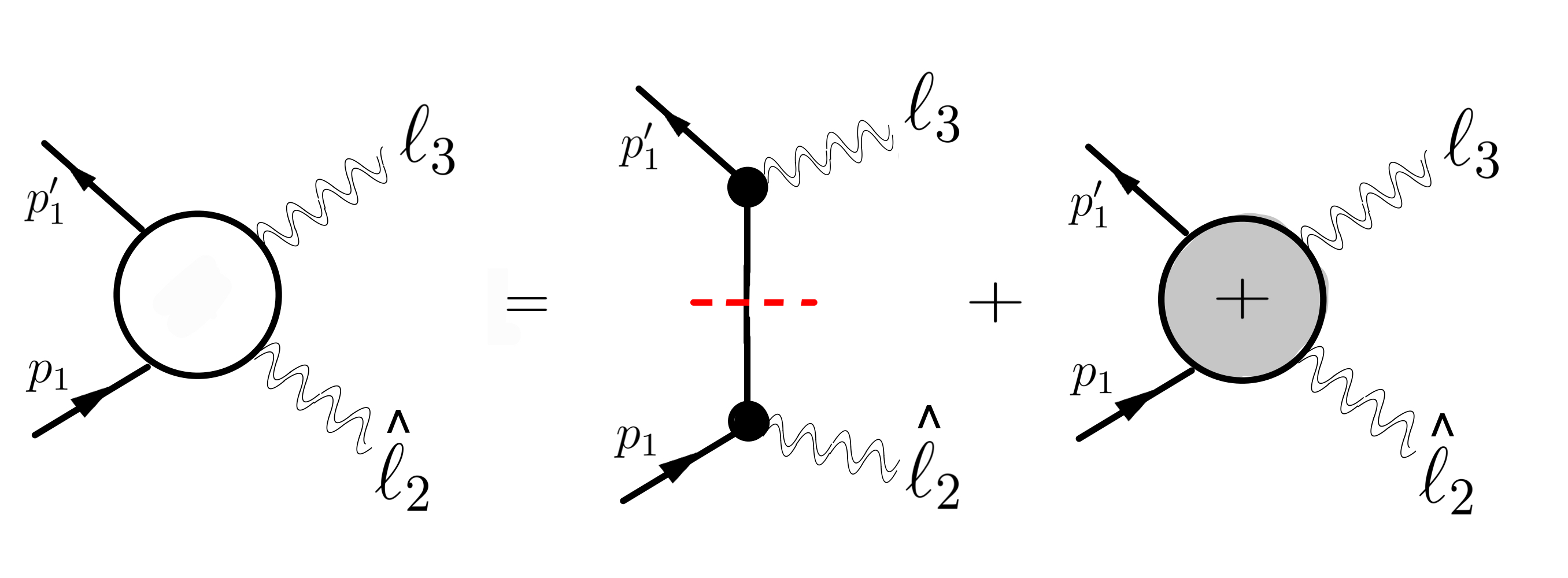}
\caption{The four-point relation graphically outlined. The red line
  symbolises a delta-function constraint, the grey blob is $M^{\textrm{tree}(+)}_2$. }
 \end{figure}
In the case the `hatted'  momentum enters the incoming momentum as in
$M^{\rm tree}_2(p_1+\hat{\ell}_4,\ell_2,\ell_3,-p_1')$ with
$\ell_4=-q+\ell_2+\ell_3$ we have
\begin{multline}\label{e:M4ext}
  M^{\rm tree}_2(p_1+\hat{\ell}_4,\ell_2,\ell_3,-p_1')=M^{\textrm{tree}(+)}_2(p_1+\hat{\ell}_4,\ell_2,\ell_3,-p_1')\cr
  +\deltabar((p_1+\hat{\ell}_4+\ell_2)^2-m_1^2)
M^{\textrm{tree}(+)}_1(p_1+\hat{\ell}_4,\ell_2,-p_1-\hat{\ell}_4-\ell_2)
M^{\textrm{tree}(+)}_1(p_1+\hat{\ell}_4+\ell_2,\ell_3,-p_1') \cr
+\deltabar((p_1+\hat{\ell}_4+\ell_3)^2-m_1^2)M^{\textrm{tree}(+)}_1(p_1+\hat{\ell}_4,\ell_3,-p_1-\hat{\ell}_4-\ell_3) M^{\textrm{tree}(+)}_1(p_1+\hat{\ell}_4+\ell_3,\ell_2,-p_1'),
\end{multline}
where we have made use of the notation $\deltabar(x) \equiv -2\pi i
\delta(x)$, and  we used that for the three-point functions
\begin{equation}
 M^{\rm tree}_1(p_1+\hat{\ell}_4+\ell_2,\ell_3,-p_1')=M^{\textrm{tree}(+)}_1(p_1+\hat{\ell}_4+\ell_2,\ell_3,-p_1')=M^{\textrm{tree}(-)}_1(p_1+\hat{\ell}_4+\ell_2,\ell_3,-p_1') .  
\end{equation}
\subsubsection{The multi-soft expansion of the four-point amplitude}\label{sec:multifour}
We consider the following multi-soft scaling of graviton legs in the context of the two-scalar-$(L+1)$-graviton tree amplitudes described in the above sections. We take
$\ell_i\to |\vec{q}| \tilde \ell_i$ with $|\vec q|\to0$, so that
\begin{equation}
 q= p-p'=-|\vec q|(\tilde \ell_2+\tilde \ell_3).
\end{equation}
For a conventional Feynman gravity amplitude (see
appendix~\ref{sec:softKLT} for a derivation) we have the universal results
\begin{equation}\label{e:limM2Feyn}
\lim_{|\vec q|\to0} M^{\rm tree}_{2}(p,|\vec q|\tilde\ell_2,|\vec q| \tilde \ell_{3},-p') \sim {1\over|\vec{q}|}\,.
\end{equation}
We know derive explicitly the soft scaling behaviour of the four-point amplitude.
The four-point graviton amplitude reads (see section~\ref{sec:CHY} for details)
\begin{equation}
M^{\rm tree}_2(p,\ell_2,\ell_3,-p')
= \frac{i {N_2}(p,2,3,-p')^2}{(p+\ell_2)^2-m^2+i\varepsilon}+\frac{i{N_2}(p,3,2,-p')^2}{(p+\ell_3)^2-m^2+i\varepsilon}+\frac{i\left({N_2}^{[2,3]}\right)^2}{q^2}\,.
\end{equation}
Performing the soft expansion of numerators (see
appendix~\ref{sec:4pt} for their expressions) we get for the amplitude
the soft expansion (with $\ell_i=|q|\,\tilde\ell_i$ and $|q|\to0$)
\begin{multline}
M^{\rm tree}_2(p,\ell_2,\hat{\ell_3},-p')=-\frac{2i(p\cdot\zeta_2)^2(p\cdot\zeta_3)^2}{|\vec
 q|}\Big(\frac{1}{p\cdot\tilde\ell_2+i\varepsilon}-\frac{1}{p\cdot\tilde\ell_2-i\varepsilon}\Big)\cr
+\frac{4i(p\cdot\tilde F_2\cdot\tilde F_3.p)^2}{(p\cdot \tilde\ell_2)^2}+\mathcal{O}(|\vec q|)\,,
\end{multline}
where we defined the field strength 
$\tilde F_i^{\mu \nu}=\tilde\ell_i^\mu \zeta_i^\nu-\tilde\ell_i^\nu \zeta_i^\mu$.
This means that we have the following soft expansion of the tree amplitudes
\begin{equation}
M^{\rm tree}_2(p,\ell_2,\hat{\ell_3},-p')=-\frac{4\pi(p\cdot\zeta_2)^2(p\cdot\zeta_3)^2 \delta(p.\tilde\ell_2)}{|\vec
 q|}+\frac{4i(p\cdot\tilde F_2\cdot\tilde F_3.p)^2}{(p\cdot \tilde\ell_2)^2}+\mathcal{O}(|\vec q|)\,,
\end{equation}
so that $M^{\rm tree}_2 \sim \mathcal O(\frac{1}{|\vec q|})$ as given in~\eqref{e:limM2Feyn}.
The same derivation for the
$M^{\textrm{tree}(+)}_{2}(p,\ell_2,\hat{\ell_3},-p')$ amplitude
in~\eqref{e:M2plus} on the other hand is
\begin{multline}
M^{\textrm{tree}(+)}_2(p,\ell_2,\hat{\ell_3},-p')=-\frac{2i(p\cdot\zeta_2)^2(p\cdot\zeta_3)^2}{|\vec
 q|}\Big(\frac{1}{p\cdot\tilde\ell_2+i\varepsilon}-\frac{1}{p\cdot\tilde\ell_2+i\varepsilon}\Big)\cr
+\frac{4i(p\cdot\tilde F_2\cdot\tilde F_3.p)^2}{(p\cdot \tilde\ell_2)^2}+\mathcal{O}(|\vec q|)\,,
\end{multline}
showing that  
\begin{equation}
M^{\textrm{tree}(+)}_{2}(p,\ell_2,\hat{\ell_3},-p')=\frac{4i(p\cdot\tilde F_2\cdot\tilde F_3.p)^2}{(p\cdot \tilde\ell_2)^2}+\mathcal{O}(|\vec
                  q|) =\mathcal O(|\vec q|^0)\,.
\end{equation}                  
Similar arguments imply that $M^{\textrm{tree}(-)}_2 \sim \mathcal O(|\vec q|^0)$.
Therefore the amplitude $M_{2}^\pm$ have the multi-soft behaviour 
\begin{equation}
\lim_{|\vec q|\to0} M^{\textrm{tree}(\pm)}_{2}(p,|\vec q|\tilde\ell_2,|\vec q|\hat{\tilde\ell}_{3},-p') \sim |\vec{q}|^0\,.
\end{equation}
This behaviour generalises to the general case
$M^{\textrm{tree}(\pm)}_{2}(p,|\vec q|\tilde\ell_2,\dots ,|\vec
q|\hat{\tilde\ell}_{L+2},-p')$.

%------------------------------------------------------------------
\subsection{The five-point case}\label{sec:fivepointtree}
For making the multi-soft behaviour  at five points,  we start by
rewriting the tree amplitude in the following  form
\begin{multline}
M^{\rm tree}_3(p_1,\ell_2,\ell_3,\hat{\ell}_4,-p_1')=
\Bigg(\frac{n_{p_1+\ell_2,p_1+\ell_2+\ell_3}}{((p_1+\ell_2)^2-m_1^2+i
  \varepsilon)((p_1+\ell_2+\ell_3)^2-m_1^2+i \varepsilon)}
\\+
\frac{n_{p_1+\ell_2,\ell_3+\hat{\ell}_4}}{((p_1+\ell_2)^2-m_1^2+i \varepsilon)(\ell_3+\hat{\ell}_4)^2}
+
\frac{n_{p_1+\ell_2+\ell_3,\ell_2+\ell_3}}{((p_1+\ell_2+\ell_3)^2-m_1^2+i \varepsilon)(\ell_2+\ell_3)^2}
\\+
\frac{n_{p_1+p_2,\ell_3+\hat{\ell}_4}}{q^2(\ell_3+\hat{\ell}_4)^2}
 \Bigg)+ {\rm perm.} \{\ell_2,\ell_3,\hat{\ell}_4\}\,,
\end{multline}
where we impose momentum conservation on $\hat
\ell_4=-q-\ell_2-\ell_3$ and the cut condition $\ell_2^2=\ell_3^2=\hat\ell_4^2=0$.
Flipping the propagators as in the four-point case, we get 
\begin{multline}
M^{\rm tree}_3(p_1,\ell_2,\ell_3,\hat{\ell}_4,-p_1')=
n_{p_1+\hat{\ell}_4,p_1+\ell_2+\hat{\ell}_4}
\deltabar((p_1+\hat{\ell}_4)^2-m_1^2)\deltabar((p_1+\ell_2+\hat{\ell}_4)^2-m_1^2)\\+
n_{p_1+\hat{\ell}_4,p_1+\ell_3+\hat{\ell}_4}
\deltabar((p_1+\hat{\ell}_4)^2-m_1^2)\deltabar((p_1+\ell_3+\hat{\ell}_4)^2-m_1^2)\cr
+\Big(\frac{
 n_{p_1+\ell_2,p_1+\ell_2+\hat{\ell}_4}}{(p_1+\ell_2)^2-m_1^2+i\varepsilon}
+ \frac{ n_{p_1+\hat{\ell}_4,p_1+\ell_2+\hat{\ell}_4} }{(p_1+\hat{\ell}_4)^2-m_1^2+i\varepsilon}+\frac{ n_{p_1+\ell_2+\hat{\ell}_4,\ell_1+\hat{\ell}_4}}{(\ell_2+\hat{\ell}_4)^2}\Big) \deltabar((p_1+\ell_2+\hat{\ell}_4)^2-m_1^2)\cr
\Big(\frac{n_{p_1+\ell_3,p_1+\ell_3+\hat{\ell}_4}
}{(p_1+\ell_3)^2-m_1^2+i\varepsilon}+\frac{ n_{p_1+\hat{\ell}_4,p_1+\ell_3+\hat{\ell}_4} }{(p_1+\hat{\ell}_4)^2-m_1^2+i\varepsilon}+\frac{n_{p_1+\ell_3+\hat{\ell}_4,\ell_3+\hat{\ell}_4} }{(\ell_3+\hat{\ell}_4)^2}\Big)  \deltabar((p_1+\ell_3+\hat{\ell}_4)^2-m_1^2)
\cr\Big(\frac{ n_{p_1+\hat{\ell}_4,p_1+\ell_2+\hat{\ell}_4} }{(p_1+\ell_2+\hat{\ell}_4)^2-m_1^2-i\varepsilon}+\frac{n_{p_1+\hat{\ell}_4,p_1+\ell_3+\hat{\ell}_4} }{(p_1+\ell_3+\hat{\ell}_4)^2-m_1^2-i\varepsilon}+\frac{ n_{p_1+\hat{\ell}_4,\ell_2+\ell_3} }{(\ell_2+\ell_3)^2}\Big) \deltabar((p_1+\hat{\ell}_4)^2-m_1^2)\cr+M^{\textrm{tree}(+)}_3(p_1,\ell_2,\ell_3,\hat{\ell}_4,-p_1').
\end{multline}
With $\deltabar(x)\equiv-2\pi i \delta(x)$.
From the factorisation property of the tree-level amplitude we know
that coefficient of the delta-function are products of tree-level
amplitudes
\begin{multline}
 n_{p_1+\hat{\ell}_4,p_1+\ell_2+\hat{\ell}_4} \Big| _{(p_1+\hat{\ell}_4)^2-m_1^2=(p_1+\ell_2+\hat{\ell}_4)^2-m_1^2=0} =\\ M^{\rm tree}_1(p_1,\hat{\ell}_4,-p_1-\hat{\ell}_4)M^{\rm tree}_1(p_1+\hat{\ell}_4,\ell_2,-p_1-\hat{\ell}_4-\ell_2)M^{\rm tree}_1(p_1+\hat{\ell}_4+\ell_2,\ell_3,-p_1'),
\end{multline}
and
\begin{multline}
n_{p_1+\hat{\ell}_4,p_1+\ell_3+\hat{\ell}_4} \Big| _{(p_1+\hat{\ell}_4)^2-m_1^2=(p_1+\ell_3+\hat{\ell}_4)^2-m_1^2=0} =\\ M^{\rm tree}_1(p_1,\hat{\ell}_4,-p_1-\hat{\ell}_4)M^{\rm tree}_1(p_1+\hat{\ell}_4,\ell_3,-p_1-\hat{\ell}_4-\ell_3)M^{\rm tree}_1(p_1+\hat{\ell}_4+\ell_3,\ell_2,-p_1'),
\end{multline}
and
\begin{multline}
\Bigg(\frac{n_{p_1+\ell_2,p_1+\ell_2+\hat{\ell}_4}}{(p_1+\ell_2)^2-m_1^2+i\varepsilon}+\frac{n_{p_1+\hat{\ell}_4,p_1+\ell_2+\hat{\ell}_4}}{(p_1+\hat{\ell}_4)^2-m_1^2+i\varepsilon}+\frac{n_{p_1+\ell_2+\hat{\ell}_4,\ell_2+\hat{\ell}_4}}{(\ell_2+\hat{\ell}_4)^2}\Bigg) \Big| _{(p_1+\ell_2+\hat{\ell}_4)^2-m_1^2=0} \cr= M^{\rm tree}_2(p_1,\ell_2,\hat{\ell}_4,-p_1-\ell_2-\hat{\ell}_4)M^{\rm tree}_1(p_1+\hat{\ell}_4+\ell_2,\ell_3,-p_1'),
\end{multline}
and
\begin{multline}
\Bigg(\frac{n_{p_1+\ell_3,p_1+\ell_3+\hat{\ell}_4}}{(p_1+\ell_3)^2-m_1^2+i\varepsilon}+\frac{n_{p_1+\hat{\ell}_4,p_1+\ell_3+\hat{\ell}_4}}{(p_1+\hat{\ell}_4)^2-m_1^2+i\varepsilon}+\frac{n_{p_1+\ell_3+\hat{\ell}_4,\ell_3+\hat{\ell}_4}}{(\ell_3+\hat{\ell}_4)^2}\Bigg) \Big| _{(p_1+\ell_3+\hat{\ell}_4)^2-m_1^2=0} \cr= M^{\rm tree}_2(p_1,\ell_3,\hat{\ell}_4,-p_1-\ell_3-\hat{\ell}_4)M^{\rm tree}_1(p_1+\hat{\ell}_4+\ell_3,\ell_2,-p_1'),
\end{multline}
and
\begin{multline}
\Bigg(\frac{n_{p_1+\hat{\ell}_4,p_1+\ell_2+\hat{\ell}_4}}{(p_1+\ell_2+\hat{\ell}_4)^2-m_1^2+i\varepsilon}+\frac{n_{p_1+\hat{\ell}_4,p_1+\ell_3+\hat{\ell}_4}}{(p_1+\ell_3+\hat{\ell}_4)^2-m_1^2+i\varepsilon}+\frac{n_{p_1+\hat{\ell}_4,\ell_2+\ell_3}}{(\ell_2+\ell_3)^2}\Bigg) \Big| _{(p_1+\hat{\ell}_4)^2-m_1^2=0} \cr =M^{\rm tree}_1(p_1,\hat{\ell}_4,-p_1-\hat{\ell}_4)M^{\rm tree}_2(p_1+\hat{\ell}_4,\ell_2,\ell_3,-p_1').
\end{multline}
The factorisation relations involve
$M^{\rm tree}_2(p_1+\hat{\ell}_4,\ell_2,\ell_3,-p_1')$ which is rewritten
using~\eqref{e:M4ext}. This leads to the following expression suitable
for the multi-soft expansion
\begin{multline}
 M^{\rm tree}_3(p_1,\ell_2,\ell_3,\hat{\ell}_4,-p_1')=
 \deltabar((p_1+\hat{\ell}_4)^2-m_1^2)\deltabar((p_1+\ell_2+\hat{\ell}_4)^2-m_1^2)\cr
 \times
 M^{\textrm{tree}(+)}_1(p_1,\hat{\ell}_4,-p_1-\hat{\ell}_4)M^{\textrm{tree}(+)}_1(p_1+\hat{\ell}_4,\ell_2,-p_1-\hat{\ell}_4-\ell_2)M^{\textrm{tree}(+)}_1(p_1+\hat{\ell}_4+\ell_2,\ell_3,-p_1')\cr
 +
 \deltabar((p_1+\hat{\ell}_4)^2-m_1^2)\deltabar((p_1+\ell_3+\hat{\ell}_4)^2-m_1^2)
 \cr
 \times
 M^{\textrm{tree}(+)}_1(p_1,\hat{\ell}_4,-p_1-\hat{\ell}_4)M^{\textrm{tree}(+)}_1(p_1+\hat{\ell}_4,\ell_3,-p_1-\hat{\ell}_4-\ell_3)M^{\textrm{tree}(+)}_1(p_1+\hat{\ell}_4+\ell_3,\ell_2,-p_1')\cr
 +\deltabar((p_1+\ell_2+\hat{\ell}_4)^2-m_1^2)
 M^{\textrm{tree}(+)}_2(p_1,\ell_2,\hat{\ell}_4,-p_1-\ell_2-\hat{\ell}_4)M^{\textrm{tree}(+)}_1(p_1+\hat{\ell}_4+\ell_2,\ell_3,-p_1')\cr
 +\deltabar((p_1+\ell_3+\hat{\ell}_4)^2-m_1^2)M^{\textrm{tree}(+)}_2(p_1,\ell_3,\hat{\ell}_4,-p_1-\ell_3-\hat{\ell}_4)M^{\textrm{tree}(+)}_1(p_1+\hat{\ell}_4+\ell_3,\ell_2,-p_1')
 \cr
 +
 \deltabar((p_1+\hat{\ell}_4)^2-m_1^2)M^{\textrm{tree}(+)}_1(p_1,\hat{\ell}_4,-p_1-\hat{\ell}_4)M^{\textrm{tree}(+)}_2(p_1+\hat{\ell}_4,\ell_2,\ell_3,-p_1')\cr
 +M^{\textrm{tree}(+)}_3(p_1,\ell_2,\ell_3,\hat{\ell}_4,-p_1'),
\end{multline}
with a similar expression involving $M^{\textrm{tree}(-)}_3$ after
flipping the $+i\varepsilon$ poles and with $\deltabar(x)\equiv-2\pi i \delta(x)$.

%------------------------------------------------------------------------
\subsection{General $(L+1)$ graviton case}
\begin{figure}[ht]
 \centering
\includegraphics[width=14cm]{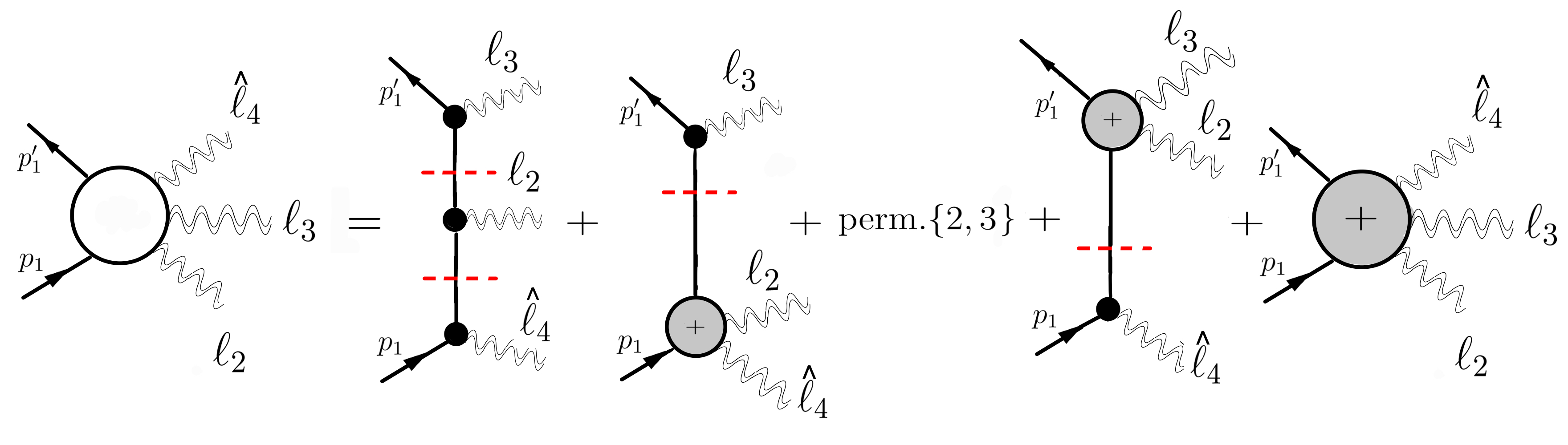}
\caption{The five-point relation graphically outline depicted. The red line symbolises a delta-function constraint. }
 \end{figure}

The symbolic structure of the four-point amplitudes derived in the
previous sections read 
\begin{equation}
M^{\rm tree}_{2} \sim (M^{\textrm{tree}(+)}_1)^2 \delta_i(\ldots) +M^{\textrm{tree}(+)}_{2}\,,
\end{equation}
and the five-point amplitudes read 
\begin{equation}
M^{\rm tree}_{3} \sim (M^{\textrm{tree}(+)}_1)^{3} \prod_i^2 \delta_i(\ldots) +M^{\textrm{tree}(+)}_1 M^{\textrm{tree}(+)}_2 \delta_i(\ldots)+M^{\textrm{tree}(+)}_{3}\,.
\end{equation}
Using the expression for the six- and seven-points amplitudes
presented in section~\ref{sec:CHY} we have derived similar expressions,
which take the symbolic form
\begin{multline}
M^{\rm tree}_{4} \sim (M^{\textrm{tree}(+)}_1)^{4} \prod_i^3
                   \delta_i(\ldots) +(M^{\textrm{tree}(+)}_1)^{2}
                   (M^{\textrm{tree}(+)}_2) \prod_i^{2}
                   \delta_i(\ldots)\cr
  +M^{\textrm{tree}(+)}_1 M^{\textrm{tree}(+)}_3
                   \delta(\ldots)+M^{\textrm{tree}(+)}_{4},
\end{multline}
\begin{multline}  
M^{\rm tree}_{5} \sim (M^{\textrm{tree}(+)}_1)^{5} \prod_i^4
\delta_i(\ldots) +(M^{\textrm{tree}(+)}_1)^{3}
(M^{\textrm{tree}(+)}_2) \prod_i^{3} \delta_i(\ldots)+ \cdots
\cr+M^{\textrm{tree}(+)}_1 M^{\textrm{tree}(+)}_4
\delta(\ldots)+M^{\textrm{tree}(+)}_{5},
\end{multline}
It follows on general grounds from the above examples that we have the
following structure for the general $L$ point case organised by powers
of unitarity cuts delta-functions
\begin{multline}\label{e:MLdelta}
M^{\rm tree}_{L+1} \sim (M^{\textrm{tree}(+)}_1)^{L+1} \prod_i^L
\delta_i(\ldots) +(M^{\textrm{tree}(+)}_1)^{L-1}
(M^{\textrm{tree}(+)}_2) \prod_i^{L-1} \delta_i(\ldots)+\cdots\cr
+M^{\textrm{tree}(+)}_1 M^{\textrm{tree}(+)}_L \delta(\ldots)+M^{\textrm{tree}(+)}_{L+1}\,,
\end{multline}
with a similar expansion involving the $M_{L+1}^{\textrm{tree}(-)}$ amplitudes.

%-----------------------------------------------------------------------
\subsection{Multi-soft graviton of the tree-level amplitudes}\label{sec:multisoft}
We consider now the following multi-soft scaling of graviton legs in the context of the two-scalar-$(L+1)$-graviton tree amplitudes described in the above sections. We take
$\ell_i\to |\vec{q}| \tilde \ell_i$ with $|\vec q|\to0$, so that
\begin{equation}
 q= p-p'=-|\vec q|\sum_{i=2}^{L+2} \tilde \ell_i.
\end{equation}
We find the following results. For a conventional Feynman gravity amplitude (see
appendix~\ref{sec:softKLT} for a derivation) we have the universal results
\begin{equation}
\lim_{|\vec q|\to0} M^{\rm tree}_{L+1}(p,|\vec q|\tilde\ell_2,\dots ,|\vec q| \tilde \ell_{L+2},-p') \sim |\vec{q}|^{-L}\,.
\end{equation}
In contrast to this, the amplitude $M_{L+1}^\pm$ have the multi-soft behaviour 
\begin{equation}\label{e:Mtildescaling}
\lim_{|\vec q|\to0} M^{\textrm{tree}(\pm)}_{L+1}(p,|\vec q|\tilde\ell_2,\dots ,|\vec q|\hat{\tilde\ell}_{L+2},-p') \sim |\vec{q}|^0\,.
\end{equation}
This is trivial for the three-point amplitude $M^{\textrm{tree}(+)}_{1}(p,|\vec
q|\hat{\tilde\ell}_2,-p')$ as such amplitudes are independent of the
graviton momentum. The case of the four-point amplitude $L=2$ has been
shown in section~\ref{sec:multifour}.
For higher-multiplicity amplitudes, this can be checked, considering the explicit expressions of the tree-level amplitudes provided up to seven points.

\subsubsection{The multi-soft expansion of the generic gravity amplitude}\label{sec:multigen}
By combining~\eqref{e:Mtildescaling} with the scaling of the
delta-function 
\begin{equation}
 \delta{\left((p_1+ \sum \ell_i)^2-m_1^2\right)} = \delta \left(2|q| p_1\cdot \sum \tilde\ell_i+\mathcal O(|q|^2)\right)={1\over|q|} \delta\left(2 p_1\cdot \sum \tilde\ell_i\right)+\mathcal O(|q|^0),
\end{equation}
we deduce that the amplitude $\mathcal M_{L+1}$ has the multi-soft
scaling 
\begin{equation}\label{e:Treemultisoft}
\lim_{|\vec q|\to0} M^{\rm tree}_{L+1}(p,|\vec q|\tilde\ell_2,\dots,|\vec
q|\hat{\tilde\ell}_{L+2},-p')
=\frac{(M^{\rm tree}_{1})^{L+1}\delta(\cdots)^{L}}{|\vec{q}|^{L}}+ \mathcal
O\left(1\over |q|^{L-1}\right)\,.
\end{equation}
%
%------------------------------------------------------------------------
\subsection{The classical part from the multi-soft graviton expansion}\label{sec:classicalmulti}
We will now explain how the above multi-soft expansions of the tree-level amplitude in
the multi-graviton cut imply a reorganisation of the computations of
integrands, and allow us to easily identify classical components in the amplitude.\\[5pt]
In the limit $\hbar\to0$, $q\to0$ with $\underline
q=q/\hbar$ fixed,  the multi-loop amplitude has the following $\hbar$-Laurent
expansion (see section~3
of~\cite{Bjerrum-Bohr:2021vuf} for details)
\begin{equation}\label{e:MLLaurent}
  \mathcal M_L(\sigma,|q|)={1\over \hbar^{L-1} |q|^{(4-D)L}}
   \sum_{r\geq -2} \mathcal M^{(r)}_L(\sigma,\epsilon) (\hbar |q|)^r.
  \end{equation}
We will now discuss how the organisation of the tree-level amplitudes across the
cut in section~\ref{sec:unitaritycuts} allows us to identify the classical part of the integrand.\\[5pt]
When plugging the delta-function expansion from~\eqref{e:MLdelta} in the product of tree-level amplitudes, the integrand of the cut
integral in~\eqref{e:MLcut}, becomes a sum of contributions
organised as follows
\begin{equation}\label{e:MLcutdelta}
 \mathcal M_{L}^{\rm cut}\sim\sum_{k=0}^{2L}  \hbar^{3L+1} \int {(d^D\ell)^L\over \hbar^{DL}}\frac{
 \left(\delta((p_1+\sum_i \ell_{\alpha_i})^2-m_1^2)\right)^{k} \times (\prod M^{\textrm{tree}(+)}) \times
 (\prod M^{\textrm{tree}(-)\,\dagger})}{(\ell^2)^{L+1}}.
\end{equation}
Now in the soft-expansion $|q|=\hbar \underline q$ with $\ell=\hbar
|\underline q|\hat \ell$ for $\hbar\to0$, the delta-function behaves at leading
order as
\begin{equation}
 \lim_{|q|\to0} \delta((p+\ell)^2-m^2)\sim \delta(2\hbar |q| p\cdot\hat\ell) \sim
 \frac{\delta(2p\cdot\hat\ell)}{\hbar|\underline q|}.
\end{equation}
The multi-soft expansion of
section~\ref{sec:multisoft} stipulates that the amplitudes $M^\pm$ are of order
$\mathcal O(|q|^0)$, so the generic integrals in the multi-graviton cut
behaves as
\begin{equation}
 \mathcal M_{L}^{\rm cut}\sim \sum_{k=0}^{2L}  {\hbar^{L-1-k}\over
 |\underline q|^{2+k-(D-2)L}}.
\end{equation}
We can have three type of contributions:
\paragraph{ Terms with $k=L$ delta-functions.} Such contributions, behave as
 \begin{equation}
     {1\over \hbar |\underline q|^{2-(D-3)L}},
    \end{equation}
    which is of classical order and given by terms with $r=L-2$ in~\eqref{e:MLLaurent}. Thus in this case,
\begin{equation}\label{e:Mclassical}
 \mathcal M_L(\sigma,|q|)\Big|_{\rm{classical}}= {1\over \hbar}
 {\mathcal M_L^{(L-2)}(\sigma,\epsilon)\over |\underline q|^{2-(D-3)L}},
\end{equation}
which implies that for computing the classical part of the amplitude,
we can approximate the unitarity delta-function constraint as a velocity
cut $\delta((p+\ell)^2-m^2)\sim \delta(2p\cdot\ell)$, which hugely simplifies 
the integral computation.
This classical term receives two kinds of contributions discussed in
section~5.2 of~\cite{Bjerrum-Bohr:2021din}: (1)
disconnected graphs and (2) connected graphs as in figure~7 and~8
of that paper. As a consequence of the unitarity relations between the
classical scattering matrix element and the radial action $\hat N$
derived in section~2 of~\cite{Damgaard:2021ipf}, such factorised contributions are
cancelled by unitarity and do not contribute to the radial action. We will see an explicit
example in section~\ref{sec:oneloopradial} below.
\paragraph{ Terms with $k<L$ delta-functions.} They  are of order $\mathcal
 O(\hbar^0)$ and correspond to quantum contributions.
 \paragraph{ Terms with $k>L$ delta-functions.} They correspond to contributions
  with $-2\leq r\leq L-3$ in the Laurent
  expansion~\eqref{e:MLLaurent}. It was shown 
  in~\cite{Bjerrum-Bohr:2021vuf,Bjerrum-Bohr:2021din} that these
  contributions to the one-loop and two-loop amplitudes exponentiate
  and do not contribute to the classical part. These products of
  unitarity delta-functions are precisely the ones arising from the expansion
  of the exponential representation of the $S$-matrix, see ref.~\cite{Damgaard:2021ipf}.

We note that the decomposition of the integrand in~\eqref{e:MLcutdelta} as a sum of
powers of unitarity cut delta-functions realizes
the expansion of the exponential representation of the $S$-matrix
given in~\cite{Damgaard:2021ipf}. The classical part of the amplitude
is the radial action $N_L$. We will now develop this into a new practical tool for computation of post-Minkowskian physics.

\subsection{The one-loop radial action}\label{sec:oneloopradial}
We will in this section illustrate how the considerations from the previous sections are useful for computations of classical contributions from scattering amplitudes. We will start by rederiving the classical contribution
from the one-loop amplitude. The two-particle cut along the two
graviton lines reads,
\begin{equation}\label{e:M1cut}
i \mathcal M_1^{\rm
 cut}(\sigma,q^2)
=\frac{1}{2}\int \frac{d^D
 \ell_3}{(2\pi)^{2D}} \frac{
 M^{\rm tree}_2(p_1,\hat{\ell}_2,\ell_3,-p_1')\times (M^{\rm tree}_2)^{\dagger}(p_2,-\hat{\ell}_2,-\ell_3,-p_2')}{\hat{\ell}_2^2 \ell_3^2},
\end{equation}
where $\hat \ell_2=-q-\ell_3$.

Using the expressions in~\eqref{e:M2plus} and~\eqref{e:M2minus} for
the tree-level amplitude in the cut, we obtain
\begin{align}
&i \mathcal M_1^{\rm cut}(\sigma,q^2)=\frac{1}{2}\int \frac{d^D
 \ell_3}{(2\pi)^{2D-2}} {\delta((p_1+\hat{\ell}_2)^2-m_1^2)
\delta((p_2-\hat{\ell}_2)^2-m_2^2) \over\hat{\ell}_2^2
 \ell_3^2} \cr \times 
 &M^{\textrm{tree}(+)}_1(p_1,\hat{\ell}_2,-p_1-\hat{\ell}_2) 
 M^{\textrm{tree}(+)}_1(p_1+\hat{\ell}_2,\ell_3,-p_1')\cr
&\times M^{\textrm{tree}(-)\,\dagger}_1(p_2,-\hat{\ell}_2,-p_2+\hat{\ell}_2)
 M^{\textrm{tree}(-)\,\dagger}_1(p_2-\hat{\ell}_2,-\ell_3,-p_2')\cr
&-\frac{i}{2}\int \frac{d^D \ell_3}{(2\pi)^{2D-1}}
{\delta((p_1+\hat{\ell}_2)^2-m_1^2)\over \hat{\ell}_2^2
 \ell_3^2}\cr
&\times M^{\textrm{tree}(+)}_1(p_1,\hat{\ell}_2,-p_1-\hat{\ell}_2) 
 M^{\textrm{tree}(+)}_1(p_1+\hat{\ell}_2,\ell_3,-p_1')M_2^{\textrm{tree}(-)\,\dagger}(p_2,-\hat{\ell}_2,-\ell_3,-p_2')\cr
&+\frac{i}{2}\int \frac{d^D \ell_3}{(2\pi)^{2D-1}}
{\delta((p_2-\hat{\ell}_2)^2-m_2^2)\over \hat{\ell}_2^2
                \ell_3^2} \cr
                &\times M^{\textrm{tree}(+)}_2(p_1,\hat{\ell}_2,\ell_3,-p_1')
 M^{\textrm{tree}(-)\,\dagger}_1(p_2,-\hat{\ell}_2,-p_2+\hat{\ell}_2)
M^{\textrm{tree}(-)\,\dagger}_1(p_2-\hat{\ell}_2,-\ell_3,-p_2')\cr
&+\frac{1}{2}\int \frac{d^D \ell_3}{(2\pi)^{2D}}\frac{
M^{\textrm{tree}(+)}_2(p_1,\hat{\ell}_2,\ell_3,-p_1') M_2^{\textrm{tree}(-)\,\dagger}(p_2,-\hat{\ell}_2,-\ell_3,-p_2')}{\hat{\ell}_2^2 \ell_3^2}.
\end{align}
In the first three lines we recognize the factorisation of the four-point
tree-level amplitudes $\mathcal M_0(p_1,p_2,-p_1-\ell_1,-p_2-\ell_1)$ and
$\mathcal M_0(p_1+\ell_1,p_2+\ell_1,-p_1',-p_2')$ on the massless graviton pole.
The contribution proportional to $M^{\textrm{tree}(+)}_2(p_1,\hat{\ell}_1,\ell_2,-p_1')\break M_2^{\textrm{tree}-\,\dagger}(p_2,-\hat{\ell}_1,-\ell_2,-p_2')$ on the last line can be
neglected as it is of quantum order $\frac{{\cal O}(|q|^{D}) \times
  {\cal O}(|q|^{0})}{{\cal O}(|q|^{4})}\sim {\cal O}(|q|^{D-4})$ which
is  ${\cal O}(\log(|q|))$ in $D=4$ dimensions. We thus have
\begin{align}\label{e:oneloopcut}
&i \mathcal M_1^{\rm
 cut}(\sigma,q^2)=\frac{i}{2}\int \frac{d^D \ell_2 d^D
 \ell_3}{(2\pi)^{2D-2}} \delta^{(D)}(\ell_2+\ell_3+q)
\delta((p_1+\ell_2)^2-m_1^2) \delta((p_2-\ell_2)^2-m_2^2)\cr &\times 
\mathcal M_0(p_1,p_2,-p_1-\ell_2,-p_2-\ell_2)
\mathcal M_0(p_1+\ell_2,p_2+\ell_2,-p_1',-p_2') \cr &
-\frac{1}{2}\int \frac{d^D
                                                      \ell_3}{(2\pi)^{2D-2}} \delta((p_1+\hat{\ell}_1)^2-m_1^2)\times\cr
  &%\hskip1.2cm
    \frac{ 
 M^{\textrm{tree}(+)}_1(p_1,\hat{\ell_2},-p_1-\hat{\ell}_1)
 M^{\textrm{tree}(+)}_1(p_1+\hat{\ell_2},\ell_3,-p_1') M_2^{\textrm{tree}(-)\,\dagger}(p_2,-\hat{\ell}_1,-\ell_3,-p_2')}{\hat{\ell}_1^2
    \ell_3^2}\cr
  &+\frac{1}{2}\int \frac{d^D \ell_3}{(2\pi)^{2D-2}}
    \delta((p_2-\hat{\ell}_1)^2-m_2^2)\times\cr
  &% \hskip1.2cm
    \frac{ M^{\textrm{tree}(+)}_1(p_1,\hat{\ell}_1,\ell_3,-p_1')
 M^{\textrm{tree}(-)\,\dagger}_1(p_2,-\hat{\ell_2},-p_2+\hat{\ell}_1)
 M^{\textrm{tree}(-)\,\dagger}_2(p_2-\hat{\ell_2},-\ell_3,-p_2')}{\hat{\ell}_1^2
 \ell_3^2}\cr
&+\mathcal O (|q|^{-2\varepsilon}).
\end{align}
This expression matches exactly the expansion in eq.~(2.16)
of ref.~\cite{Damgaard:2021ipf}.
Using the unitarity relation in eq.~(2.10) of~\cite{Damgaard:2021ipf},
we identify the first two lines as the product of tree-level amplitudes from
unitarity, while the rest can be associated with the one-loop contribution
to the radial action (which is the classical eikonal exponent $N_1$).

\begin{multline}\label{e:N1}
N_{1}(p_1,p_2,-p_1',-p_2')=-\frac{1}{2}\int \frac{d^D
  \ell_2}{(2\pi)^{2D-2 }} \delta((p_1+\hat{\ell}_1)^2-m_1^2)\cr\times
%&\hskip1.2cm
\frac{ 
 M^{\textrm{tree}(+)}_1(p_1,\hat{\ell_1},-p_1-\hat{\ell}_1)
 M^{\textrm{tree}(+)}_1(p_1+\hat{\ell_1},\ell_2,-p_1') M_2^{\textrm{tree}(-)\,\dagger}(p_2,-\hat{\ell}_1,-\ell_2,-p_2')}{\hat{\ell}_1^2
 \ell_2^2}\cr
+\frac{1}{2}\int \frac{d^D \ell_2}{(2\pi)^{2D-2}}
\delta((p_2-\hat{\ell}_1)^2-m_2^2)\cr
% \hskip1.2cm
\times\frac{M^{\textrm{tree}(+)}_2(p_1,\hat{\ell}_1,\ell_2,-p_1')
M^{\textrm{tree}(-)\,\dagger}_1(p_2,-\hat{\ell_1},-p_2+\hat{\ell}_1)
M^{\textrm{tree}(-)\,\dagger}_1(p_2-\hat{\ell_1},-\ell_2,-p_2')}{\hat{\ell}_1^2
\ell_2^2}\cr
+\mathcal O (|q|^{-2\varepsilon}).
\end{multline}

%----------------------------------------------------------------------
\section{Probe radial action }\label{sec:probeamp}
\begin{figure}[h]
 \centering
 \includegraphics[width=7cm]{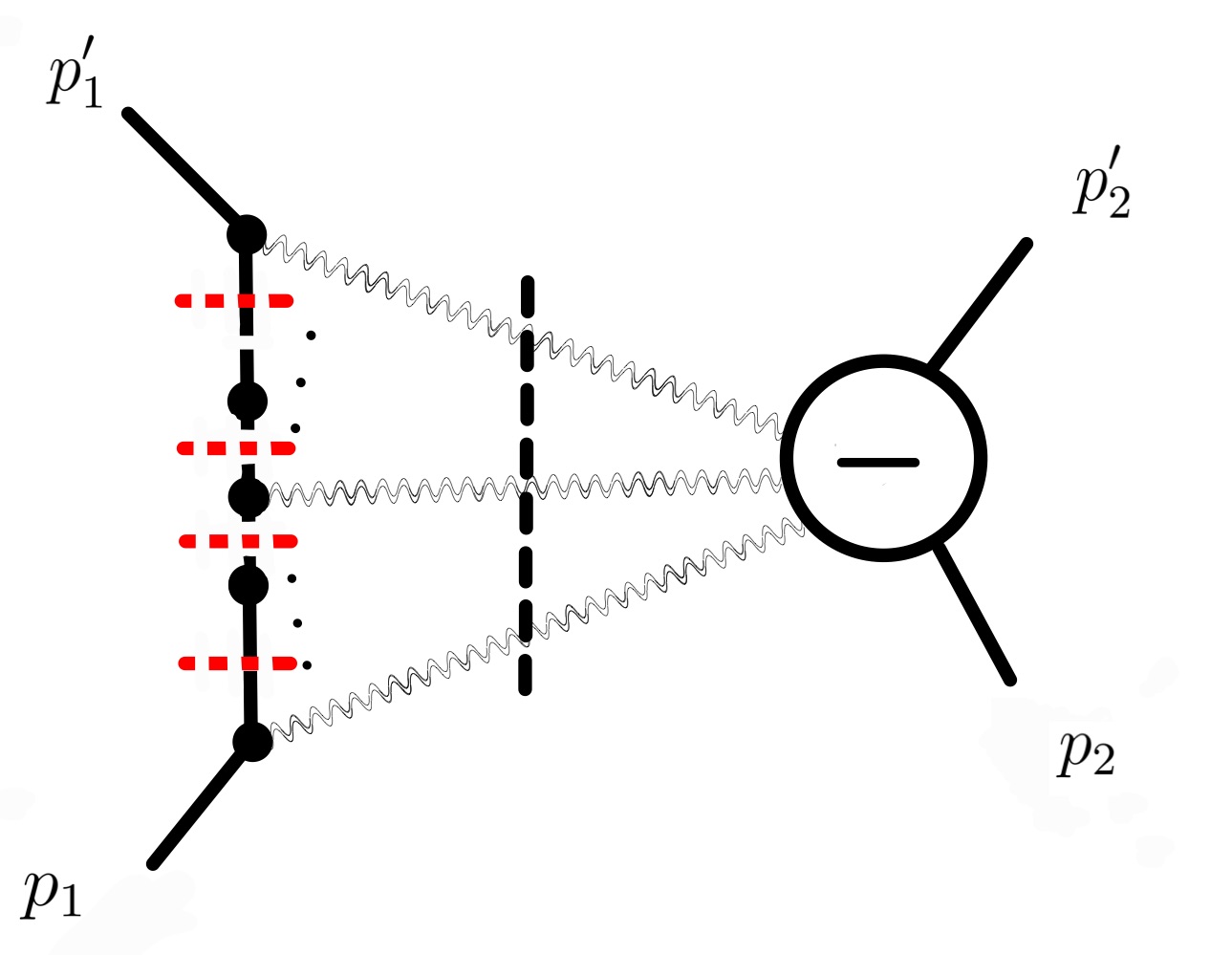}
\caption{The contribution with $n=1$ to compute the probe
  amplitude. The heavy mass $m_1$ is on the left and the small mass
  $m_2$ is on the right side.}\label{fig:probe} 
 \end{figure}

\begin{figure}[ht]
 \centering
 \includegraphics[width=8cm]{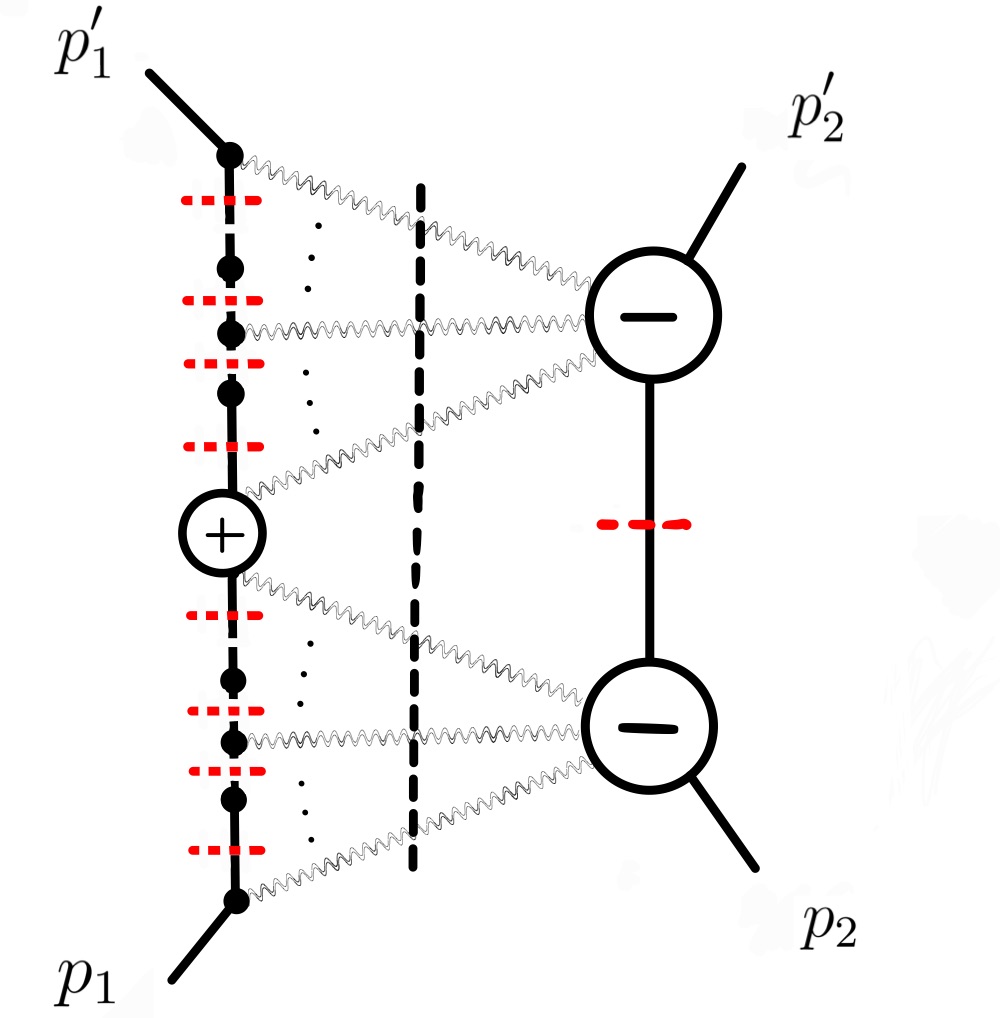}
\caption{An example of an integrand contribution at the next-to-probe
  order.  The heavy mass $m_1$ is on the left and the small mass
  $m_2$ is on the right side.} \label{fig:nextprobe}
 \end{figure}

We will now see how the  organisation of the integrand of the
multi-graviton cut integral gives a direct identification of 
the  integrand of the
classical part at $L$-loop in the probe limit $m_1\gg m_2$.
This follows from the discussion in 
section~\ref{sec:classicalmulti} using that the classical part of the
multi-loop amplitude have the symbolic representation
\begin{multline}\label{e:probeexpand}
 \mathcal M_L^{\rm classical}\sim\int 
\sum_{n=1}^{L+1}   \left(\delta((p+\sum_i \ell_i)^2-m^2)\right)^{L} \times ( M^{\textrm{tree}(+)})^{L+2-n} \times (
   M^{\textrm{tree}(-)\,\dagger})^{n}
\cr\times   \delta(q+\sum_{i=1}^{L+1}\ell_i)
   \prod_{i=1}^{L+1} {d^D\ell_i\over
   \ell_i^2}\,.
\end{multline}
We see that the leading  probe contribution arises from the 
term with $n=1$ in the integrand of the
multi-graviton cut in~\eqref{e:MLcut} 
\begin{multline}
(M_{\rm Left} {M_{\rm Right}}^\dagger)\Big|_{\rm
  probe}=
M^{\textrm{tree}(-)\dagger}_{L+1}(p_2,-\hat{\ell}_2,-\ell_3,\ldots,-\ell_{L+1},-p_2')\cr
\times M^{\textrm{tree}(+)}_1(p_1,\hat{\ell}_2,\!-p_{1}\!-\!\hat{\ell}_2)\prod_{j=3}^{L}\delta((p_1+\hat{\ell_2}+
\cdots +\ell_{j-1})^2-m_1^2)\cr
M^{\textrm{tree}(+)}_1(p_{1}+\hat{\ell}_2+ \cdots
+\ell_{j-1},\ell_j,\!-\!p_{1}\!-\!\hat{\ell}_2- \cdots +\ell_j),
\end{multline}
evaluated on the cut $\ell_i^2=0$ for $1\leq i\leq L+1$.
This contribution is represented in figure~\ref{fig:probe}.
The next-to-probe, contribution arises from the terms with $n=2$ 
which is  represented in figure~\ref{fig:nextprobe}. The $m$th next-to-probe limit is the sum of the contributions with
$n=m+1$.\\[5pt]
\noindent
 Beyond the probe regime we need to include other contributions from
 the soft expansion, as well as other unitarity cuts (like the cut containing
self-energy contributions). Including these contributions is beyond the scope of the
 present paper, but we remark that the multiple soft expansion of the
 tree amplitude lead to an efficient organisation of the integrand in such cases as well.
\subsection{Probe action from the multi-soft limits}
Our first practical computational example of  the formalism presented will be of
the computation of the probe limit of the classical amplitude at various
post-Minkowskian orders. In the probe approximation with $m_1\gg m_2$,
we consider the contribution with $n=1$
in~\eqref{e:probeexpand} which is represented on
figure~\ref{fig:probe}.
We begin at the second post-Minkowskian order where the part of the
integrand $N_{1}$ in~\eqref{e:N1}, which correspond to the probe mass being $m_2$ is,
\begin{multline}\label{probeN1}
 N_1^{\rm probe}(\sigma,\vec q)=-\frac{(32\pi G_N)^2}{2}\int
 {d^D\ell_1d^D\ell_2\over(2\pi)^{2D-2}}{\delta((p_1+\ell_1)^2-m_1^2)\over
  \ell_1^2\ell_2^2}\delta(\ell_1+\ell_2+q)\cr
 \times\sum_{h_1,h_2=\pm2}
 M^{\textrm{tree}(+)}_1(p_1,{\ell^{h_1}_1},-p_1-\ell^{h_1}_1)
 M^{\textrm{tree}(+)}_1(p_1+\ell_1^{h_1},\ell_2^{h_2},-p_1')\cr\times M_2^{\textrm{tree}(-)\,\dagger}(p_2,-\ell^{h_1}_1,-\ell^{h_2}_2,-p_2').
\end{multline}
We know that the three-point tree-level amplitudes are given by 
\begin{align}
 M^{\textrm{tree}(+)}_1(p_1,{\ell_1},-p_1-\ell_1) &= 2 p_1^\mu p_1^\nu\zeta_{1\,\mu\nu},\cr
 M^{\textrm{tree}(+)}_1(p_1+\ell_1,\ell_2,-p_1')   &=   2 (p_1^\mu+\ell_1^\mu) (p_1^\nu+\ell_1^\nu)\zeta_{2\,\mu\nu}.
\end{align}
Now to carry out the polarisation sum for the intermediate gravitons in eq. \eqref{probeN1} we use the completeness identity
\begin{equation}
\sum_{h=\pm2} \zeta^{h}_{\mu\nu} (\zeta^{h}_{\alpha\beta})^* =\frac{1}{2}\Big(\eta_{\mu\alpha}\eta_{\nu\beta}+\eta_{\mu\beta}\eta_{\nu\alpha}-\frac{2}{D-2} \eta_{\mu\nu}\eta_{\alpha\beta} \Big)\,,
\end{equation}
where we regulate the non-transverse polarization degrees of freedom by taking,
\begin{equation}
 \eta_{\mu\nu}\eta^{\mu\nu} = D-2+D_s\,,
\end{equation}
and adjust the state counting parameter $D_s$ to $D_s = 2$ to remove the dilaton~\cite{Bern:2012gh}.
In the soft limit we can evaluate the leading soft term of ${\cal O}(|q|^0)$ of $M_2^{\textrm{tree}(-)\,\dagger}$ employing that
\begin{equation}
M_2^{\textrm{tree}(-)\,\dagger}(p_2,-\ell_1,-\ell_1-q,-p_2')|_{i\varepsilon
  \to 0, \rm soft} = M^{\rm tree\,\dagger}_2(p_2,-\ell_1,-\ell_1-q,-p_2')|_{i\varepsilon \to 0, \rm soft}.
\end{equation}
Thus using the expression for $M_2^{\rm tree}$ given in~\eqref{e:M4pt}, we obtain the
following expression of the amplitude\footnote{Up to three-loop order, we can directly use the expression of the amplitude derived in the previous section. At higher-loop order, the size of integrand growths, and it is preferable to use the representation in~\eqref{e:MtreeKLTCHY} and integrate each ordering since it leads to a more convenient generation of the integrand that is easier to automate. }
\begin{multline}
 N_1^{\rm probe}(\sigma,\vec q)=2(8\pi G_N)^2\,\Big(\frac{m_2^4 m_1^4 G_N^2 ( (D-2)
  \sigma^2-1)^2}{2( D-2)^2}\mathcal I_1(2,1,1)\cr+ \frac{2m_1^4 m_2^2 G_N^2 ( (D-2)
  \sigma^2-1)}{( D-2)^2}\mathcal
  I_1(0,1,1)
 +\frac{2m_1^4G_N^2(D-3)}{(D-2)}\mathcal I_1(-2,1,1)\Big).
\end{multline}
Where we have introduced a family of integrals on which the $L$-loop
probe amplitude are expanded
\begin{multline}
 \mathcal I_L(\{a_j\},\{b_{j,k}\},\{c_j\})\equiv\cr
 \int \prod_{j=1}^{L}\frac{d^D \ell_j \delta(2p_1 \cdot \ell_j)}{(2\pi)^{D-1}}\frac{1}{\Big(2p_2 \cdot (\sum_{k=1}^{j}\ell_k)\Big)^{a_j}\prod_{k=j}^{L}\Big((\sum_{r=j}^{k}\ell_r)^2\Big)^{b_{j,k}}\Big((q+\sum_{k=1}^{j}\ell_k)^2\Big)^{c_j}}.
\end{multline}
This basis involves $3L+\frac{L(L+1)}{2}$ different variables
(including the ones in the delta functions).
 Using {\tt
 LiteRed}~\cite{Lee:2013mka} we find that only one master
integral contributes to the classical result (we have checked this to
four-loop order):
\begin{equation}
I_L(\vec{q})=\int \frac{d^D \ell_1 \cdots d^D \ell_L}{(2\pi)^{L(D-1)}}\frac{\delta(2p_1\cdot\ell_1) \cdots \delta(2p_1\cdot\ell_L)}{\ell_1^2 \cdots \ell_L^2 (\ell_1+\ell_2+ \cdots +\ell_L+q)^2}\,,
\end{equation}
where $p_1^2=m_1^2$, $p_1\cdot q=0$ and $q=(0,\vec{q})$. 
Evaluating the delta functions, the integral can be put in Euclidean form
\begin{equation}\label{e:Itriangle}
I_L(\vec{q})=\frac{(-1)^{L+1}}{2^L m_1^L}\int \frac{d^{D-1} \vec{\ell}_1 \cdots d^{D-1} \vec{\ell}_L}{(2\pi)^{L(D-1)}}\frac{1}{\vec{\ell}_1^2 \cdots \vec{\ell}_L^2 (\vec{\ell}_1+\vec{\ell}_2+ \cdots +\vec{\ell}_L+\vec{q})^2}\,,
\end{equation}
and becomes a $I_L(\vec{q})$ is a massless $L$-loop
sunset integral. This master integral also arises in the
metric computation of ref.~\cite{Mougiakakos:2020laz}. The fact that the
same master integral arises in the two computations is natural because
the probe mass $m_2$ is evolving the Schwarzschild metric sourced by
the mass $m_1$.
The master in~\eqref{e:Itriangle} is readily evaluated using the
method of section~2.2 of~\cite{Mougiakakos:2020laz}, and remarking
that $I_L(\vec{q})=(-1)^{L+1} J_{(L)}(\vec q^2)/(2^L\vec q^2m _1^L)$,
we have the result
\begin{equation}
I_L(\vec{q})=\frac{1}{(\vec{q}^2)^{1-\frac{(D-3)L}{2}}}\times
\frac{(-1)^L}{2^{L+1} m_1^{L}} \frac{\Gamma(\frac{D-3}{2})^{L}\Gamma(1-\frac{(D-3)L}{2})}{(4\pi)^{L\frac{D-1}{2}}\Gamma(\frac{(D-3)(L+1)}{2})}\,.
\end{equation}
After reduction with {\tt LiteRed}~\cite{Lee:2013mka}, the probe radial
action at $L$-loop order is
\begin{equation}\label{e:Nprobe}
 N_L^{\rm probe}(\sigma,\vec
 q)={(-1)^{L+1}4\over
   (L+1) }{(D-3)^L\over
  (D-2)^{L+1}} {c_L(\sigma,D)\over (\sigma^2-1)^L} I_L(\vec q)  \,m_1^{2(L+1)}m_2^2 (8 \pi G_N)^{L+1}  .
\end{equation}
At tree-level we have
 \begin{equation}
 c_0(\sigma,D)=\sigma ^2 (D-2)-1.
 \end{equation}
Performing the integral reduction with {\tt LiteRed}~\cite{Lee:2013mka} we find at
one-loop order
\begin{equation}
 \label{e:c1}
 c_1(\sigma,D)= \sigma ^4 (2 D-5) (2 D-3)+\sigma ^2 (30-12 D)+3,
\end{equation}
at two-loop order
\begin{equation}
 \label{e:c2}
 c_2(\sigma,D)= 2 \sigma ^6 (D-2) (3 D-8) (3 D-4)-30 \sigma ^4 (D-2) (3 D-8)+30 \sigma ^2 (3 D-8)-1,
\end{equation}
at three-loop order
\begin{multline}
 \label{e:c3}
 c_3(\sigma,D)=\frac{1}{3} \sigma ^8 (4 D-11) (4 D-9) (4 D-7) (4 D-5)-\frac{28}{3} \sigma ^6 (4 D-11) (4
         D-9) (4 D-7)\cr
         +70 \sigma ^4 (4 D-11) (4 D-9)+\sigma ^2 (1540-560
          D)+35,
\end{multline}
and finally four-loop order
\begin{multline}
 \label{e:c4}
 c_4(\sigma,D)=\frac{2}{3} \sigma ^{10} (D-2) (5 D-14) (5 D-12) (5
        D-8) (5 D-6)\cr
        -30 \sigma ^8 (D-2) (5
         D-14) (5 D-12) (5 D-8)\cr
         +420 \sigma ^6 (D-2) (5 D-14) (5 D-12)-420 \sigma ^4 (5 D-14) (5
          D-12)\cr
          +630 \sigma ^2 (5 D-14)-126.
\end{multline}
We note that the dimension dependence of the probe coefficients are compatible with 
the generic form is given in section~9 of~\cite{Brandhuber:2021eyq}.\\[5pt]
We further remark that in $D=4$ dimensions the coefficients take the simpler
factorised form
\begin{align}
 c_1(\sigma,4)&=3 (\sigma^2 -1) \left(5 \sigma ^2-1\right),\cr
c_2(\sigma,4)&=2 \left(64 \sigma ^6-120\sigma ^4+60 \sigma ^2-5\right),\cr
 c_3(\sigma,4)&=35 (\sigma^2 -1)^2 \left(33 \sigma ^4-18 \sigma^2+1\right),\cr
c_4(\sigma,4)&=6 \left(1792 \sigma ^{10}-5760 \sigma ^8+6720 \sigma ^6-3360 \sigma ^4+630 \sigma
         ^2-21\right)\,.
\end{align}
These results are in agreement with the probe limit of the two-body scattering amplitude up
to fifth post-Minkowskian order in four dimensions derived
in~\cite{Bjerrum-Bohr:2021din,Bern:2021dqo}.

\subsection{The probe amplitude from  geodesic scattering}
In this section, we will discuss how the results derived from scattering amplitudes above can be put into the context of geodesic scattering. We derive the probe amplitude from the geodesic scattering using the
Schwarzschild-Tangherlini metric in $D$ dimensions, using the
effective-one-body (EOB) formalism of
the recent ref.~\cite{Damgaard:2021rnk}. We will show that the scattering
amplitudes obtained this way matches the ones obtained from unitarity.\\[5pt]
The Schwarzschild-Tangherlini metric reads in an isotropic coordinate
system\footnote{The metric in spherical coordinates is given by ref. ~\cite{Myers:1986un}
\begin{equation}
 ds^2=\left(1-{8\pi G_N m_1 \Gamma(D-1)\over (D-2) r^{D-3}} \right) dt^2 - \left(1-{8\pi G_N m_1 \Gamma(D-1)\over (D-2) r^{D-3}}\right)^{-1} dr^2- r^2d^2\Omega_{D-2}.
\end{equation}
}
\begin{equation}
 ds^2=A(r) dt^2 + B(r) \left( dr^2+ r^2d^2\Omega_{d-2}\right)\,,
\end{equation}
where $d^2\Omega_{d-2}$ is the metric on the $D-1$ unit sphere where
\begin{equation}
 A(r)=\left(\frac{1-\frac{2 \pi ^{\frac{3-D}{2}} G_N m_1 r^{3-D} \Gamma
 \left(\frac{D-1}{2}\right)}{D-2}}{1+\frac{2 \pi ^{\frac{3-D}{2}} G_N m_1
 r^{3-D} \Gamma \left(\frac{D-1}{2}\right)}{D-2}}\right)^2,\quad
 B(r)=\left(1+\frac{2 \pi ^{\frac{3-D}{2}} G_N m_1 r^{3-D} \Gamma
 \left(\frac{D-1}{2}\right)}{D-2}\right)^{\frac{4}{D-3}}.
\end{equation}
Using the equation~(29) of~\cite{Damgaard:2021rnk} we deduce the
effective potential
\begin{equation}\label{e:Veffdef}
{ V_{\rm eff}(r)\over p_\infty^2}=1-{B(r)\over \sigma^2-1}\left({\sigma^2 \over A(r)}-1\right),
\end{equation}
with $p_\infty^2=m_2^2 (\sigma^2-1)+O(m_2/m_1)$ in the probe limit
for $m_2\ll m_1$.
Using this we deduce the scattering angle using the expression given
in~\cite{Bjerrum-Bohr:2019kec}
\begin{equation}\label{e:chidef}
 \chi^{\rm probe}(\sigma,D)= \sum_{k\geq1} {2b\over k!} \int_0^\infty du \left(d\over
 du^2\right)^k \left[{1\over u^2+b^2}\left( V_{\rm eff}\left(\sqrt{u^2+b^2}\right) (u^2+b^2)\over p_\infty^2\right)^k \right].
\end{equation}
Using that
\begin{equation}
 \int_0^\infty { b du\over (u^2+b^2)^{1+{n(D-3)\over2}}}=
 {\sqrt\pi \Gamma\left(n(D-3)+1\over2\right)
 \over 2b^{n(D-3)} \Gamma\left(n(D-3)+2\over2\right)},
 \end{equation}
 the scattering angle has the post-Minkowskian expansion in $D$
 dimensions\footnote{The expression in~\eqref{e:chiD} reproduces the
  results given in the appendix~B of~\cite{KoemansCollado:2019ggb}
to the third Post-Minkowskian order $n=3$. This was guaranteed because the
effective potential $V_{\rm eff}(r)$ in~\eqref{e:Veffdef} is
designed to  
match the scattering angle~\cite{Damgaard:2021rnk}. We remark
that the expression in~\eqref{e:chidef} for the scattering angle leads
to a much simpler evaluation of the post-Minkowskian expansion than the
derivation of the angle obtained by solving Einstein's geodesic equations as in~\cite[\S101]{Landau:1982dva} and
in~\cite{KoemansCollado:2019ggb}.}
 \begin{equation}\label{e:chiD}
 \chi^{\rm probe}(\sigma,D)= \sum_{L\geq0} {
 c_{L}(\sigma,D) \over
 (\sigma^2-1)^{L+1}} \frac{2^{L+2} \Gamma
 \left(D-1\over 2\right)^{L+1} \Gamma
 \left((D-3) (L+1)+1\over2\right)}{ (D-2)^{L+1}\pi ^{\frac{(D-3) (L+1)-1}{2} } \Gamma \left((D-3) (L+1)+2\over2\right)}\left( G_N m_1\over b^{D-3}\right)^{L+1}.
\end{equation}
We can compare with the scattering angle derived from the scattering
amplitude in the previous sections.
The Fourier transform of the probe radial action
 in~\eqref{e:Nprobe} to $b$-space
\begin{equation}
 N_L^{\rm probe}(\sigma,b)= {1\over 4m_1m_2\sqrt{\sigma^2-1}}\int{ d^{D-2}\vec q \over (2\pi)^{D-2}}
 N_L^{\rm probe}(\sigma,\vec q) e^{i\vec q\cdot \vec b}.
\end{equation}
Using the Fourier tranform of the master integral
\begin{align}
 \tilde{I}_L(\vec{b})&=\int_{\mathbb R^{D-2}} {d^{D-2}\vec q\over (2\pi)^{D-2}}   \tilde{I}_L(\vec{q}) e^{i\vec q\cdot \vec b}\cr
& =\frac{(-1)^{L-1}}{2^{2+3L} m_1^L}
  \frac{\Gamma(\frac{D-3}{2})^{L+1}\Gamma\left((D-3)(L+1)-1\over 2\right)}{\pi^{-\frac{1}{2}+(L+1)\frac{D-1}{2}}\Gamma(\frac{(D-3)(L+1)}{2})}
\Big(\frac{1}{{\vec{b}}^2}\Big)^{-\frac{1}{2}+\frac{(D-3)(L+1)}{2}}\,,
\end{align}
we obtain
\begin{multline}
 N_L^{\rm probe}(\sigma,b)= m_2
 \sqrt{\sigma^2-1} \, b\, {2^{1+L}\over\pi ^{-1+(D-3)(L+1)\over2}}\cr
 \times  {c_L(\sigma,D)
 \over (\sigma^2-1)^{L+1}} \,\frac{ \Gamma \left(\frac{D-1}{2}\right)^{L+1} \Gamma
 \left((D-3) (L+1)-1\over2\right)}{ (D-2)^{L+1} \Gamma \left((D-3) (L+1)+2\over2\right)}\left( G_N m_1\over b^{D-3}\right)^{L+1} .
\end{multline}
In the probe limit $m_2\ll m_1$ the amplitude is related to the
scattering angle by the linear relation
\begin{equation}
 \chi^{\rm probe}(\sigma,D) =-{1\over m_2\sqrt{\sigma^2-1}}
 {\partial N^{\rm probe} (b,\sigma)\over \partial b}\,,
\end{equation}
which leads to the angle at the $L+1$ post-Minkowskian order
in perfect agreement with the geodesic computation in~\eqref{e:chiD}.
%%%%%%%%%%%%%%%%%%%%%%%%%%%%%%%%%%%%%%%%%%%%%%%%%%%%%%%%%%%%%%%%%
\section{Conclusion}\label{sec:conclusion}
The scattering amplitude approach to the gravitational two-body
scattering is a promising avenue for performing post-Minkowskian
calculations needed for the construction of wave-forms and have already led 
to a renewed understanding of the connection
between quantum scattering amplitudes and classical
observables~\cite{Neill:2013wsa,Bjerrum-Bohr:2021vuf,Iwasaki:1971vb,Holstein:2004dn,Kosower:2018adc}.
Quantum scattering amplitudes contain much
more information than their classical parts and thus extracting classical
physics from amplitudes becomes more and more challenging 
at each perturbative order.\\[5pt]
In this work, combining unitarity and the concept of velocity cuts
introduced in~\cite{Bjerrum-Bohr:2021din}, we have identified exactly
those elements of integrands that lead to classical physics after
integration. Our approach uses an organisation of the integrand of
the multi-loop amplitude with unitarity cuts on the massive
scalar propagator lines used together with detailed knowledge of 
the correspondence between the multi-soft graviton
expansion and $\hbar\to0$ limit classical integrand matching 
the exponential representation of the
$S$-matrix of~\cite{Damgaard:2021ipf}.
In the classical limit, this
approach systematically relates the classical part in the scattering
amplitude to the matrix elements of the eikonal operator $\hat N$,
without having to perform the subtractions needed for the
exponentiation of the radial action.\\[5pt]
We have exemplified our approach by computing the probe amplitude at
second, third, fourth, and fifth post-Minkowskian orders. These scattering amplitudes
are obtained in the $D$-dimension. We have verified the agreement with
the results obtained by geodesic scattering in the $D$-dimensional
Schwarzschild-Tangherlini metric. \\[5pt]
We would like to emphasize that our framework for computation is not restricted to the probe integrands, but can be applied for deriving the complete post-Minkowskian scattering
potential. In ref. \cite{Brandhuber:2021eyq}, a heavy mass expansion was used to extract post-Minkowskian physics from amplitudes with applications for computing the probe limits. Although colour-kinematic numerators are also used, numerators in ref. \cite{Brandhuber:2021eyq} are different from the ones used here. It would be interesting to investigate further the connection between approaches. \\[5pt]
We also note that the nature of the external lines plays a very little role in the multi-soft scaling and localisation arguments we make -- as expected from the universal behaviour of gravitational interactions \cite{Bjerrum-Bohr:2013bxa,Bern:2020gjj,DiVecchia:2020ymx}. For instance, classical integrands in the post-Minkowskian framework with elementary spinning external particles should be possible to simplify as well using the presented formalism. Although it is interesting to study this question further we will leave it for future research work.

\acknowledgments
We thank Poul H. Damgaard for discussions and comments on the present
work. Johannes Agerskov for sharing his Mathematica package. The research of P.V. has received funding from the ANR
grant ``Amplitudes'' ANR-17- CE31-0001-01, and the ANR grant ``SMAGP''
ANR-20-CE40-0026-01 and is partially supported by the Laboratory of Mirror Symmetry NRU HSE, RF Government grant, ag. No 14.641.31.0001. P.V. is grateful to the I.H.E.S. for the use of their computer resources. The work of N.E.J.B.-B. was supported in part by the Carlsberg Foundation and by DFF grant 1026-00077B.

\appendix

%-----------------------------------------------------------------------
\section{Soft scaling from momentum kernel}\label{sec:softKLT}
%-----------------------------------------------------------------------
We will here outline the soft scaling behaviour at generic multiplicity using the momentum kernel given in eq. (2.20) of~\cite{Bjerrum-Bohr:2010pnr}.
\begin{multline}
M^{\rm tree}_{L}(p_1,\ell_2,\dots,\ell_{L+1},-p_1')=  (-1)^{L-1} \sum_{\sigma,
  \gamma\in\mathfrak S_{L}} {\mathcal S}(\sigma,\gamma)_{\ell_{L+1}}
\cr\times A_{L}(p_1,\sigma(2, \dots ,L),L+1,-p_1') A_{L}(p_1,L+1,\gamma(2, \dots ,L),-p_1')\,,
\end{multline}
where
\begin{equation}
  {\mathcal S}(\sigma,\gamma)_{\ell_{L+1}}\equiv i\prod_{t=2}^{L}
  \big(2\ell_{\gamma(t)}\cdot \ell_{ (L+1)}+\sum_{q<t} \theta(\gamma(q),\gamma(t))
  2\ell_{\gamma(t) }\cdot\ell_{\gamma(q)} \big),
  \end{equation}
  is the momentum kernel and where $\theta(i_t,i_q)$ equals 1 if the
  ordering of the legs $i_t$ and $i_q$ is opposite in the sets
  $\{i_1,\dots,i_k\}$ and $\{j_1,\dots,j_k\}$, and 0 if the ordering is
    the same. This representation is convenient as no massive momenta enter the momentum kernel. 
    
   The   colour-ordered Yang-Mills amplitudes are  $A_L$. Following the
   flipping convention, we take $\hat{\ell}_{L+1}=-q-\sum_{j=2}^{L}
   \ell_j$, and know using the arguments of
   section~\ref{sec:unitaritycuts} that the  colour-ordered amplitudes  (we have only $p_1\cdot k+i \epsilon$ propagators) satisfy
\begin{equation}
A_{L}(p_1,\sigma(2, \dots ,L),\widehat{L\!+\!1},-p_1')=A^+_{L}(p_1,\sigma(2, \dots ,L),\widehat{L\!+\!1},-p_1')\,,
\end{equation}
while
\begin{multline}
A_{L}(p_1,\widehat{L\!+\!1},\sigma(2, \dots ,L),-p_1')\!=\!\!\sum_{k=1}^{L-1} \Big(\prod_{\sum_{j=1}^{k+1}  i_j=L} \!A_{i_j}^+ \Big) (\delta( \dots ))^k\cr+A_{L}^+(p_1,\widehat{L\!+\!1},\sigma(2, \dots ,L),-p_1'),
\end{multline}
where the product in the sum contains $k$  tree-level  amplitudes
$A^+_i$. Hence we can write the generic $M^{\rm tree}_L$ amplitude in the form
\begin{multline}\label{e:MtreeKLT}
    M^{\rm tree}_{L}(p_1,\ell_2, \dots
,\widehat{\ell_{L+1}},-p_1')=M^{\textrm{tree}(+)}_{L}(p_1,\ell_2, \dots
,\widehat{\ell_{L+1}},-p_1') \cr
-(-1)^{L}\!\!\!\sum_{\sigma,
  \gamma\in\mathfrak S_{L-1}} \!\!\!  {\mathcal
  S}(\sigma,\gamma)_{\ell_{L\!+\!1}} \left(\!\!\sum_{k=1}^{L-1} (\delta( \dots ))^k\! \prod_{\sum_{j=1}^{k+1}  i_j=L} \!A_{i_j}^+ \right)  A^+_{L}(p_1,\sigma(2, \dots ,L),\widehat{L+1},-p_1') \,,
\end{multline}
where
\begin{multline}
M^{\textrm{tree}(+)}_{L}(p_1,\ell_2, \dots
,\hat{\ell}_{L+1},-p_1')=(-1)^{L-1}\sum_{\sigma, \gamma\in\mathfrak
  S_{L-1}} \mathcal S(\sigma,\gamma)_{\ell_{L+1}} \cr\times
A^+_{L}(p_1,\sigma(2, \dots ,L),\widehat{L+1},-p_1')  A^+_{L}(p_1,\widehat{L+1},\gamma(2, \dots ,L),-p_1')\,.
\end{multline}
 Now considering the expression in~\eqref{e:MtreeKLT} and picking the term
\begin{equation}
A_{L}(p_1,\widehat{L+1},L, \dots ,2,-p_1') \times A_{L}(p_1,2, \dots ,L,\widehat{L+1},-p_1')\,,
\end{equation}
we have propagator products such as
\begin{equation}
\frac{1}{(p_1+\hat{\ell}_{L+1}+\ell_L+ \dots +\ell_3)^2-m_1^2+i \epsilon}\times\frac{1}{(p_1+\ell_2)^2-m_1^2+i \epsilon}\,.
\end{equation}
After flipping we have contributions with delta functions
\begin{equation}
-\frac{2\pi i\delta((p_1-\ell_2-q)^2-m_1^2)}{(p_1+\ell_2)^2-m_1^2+i \epsilon}+\frac{1}{(p_1-\ell_2-q)^2-m_1^2+i \epsilon}\frac{1}{(p_1+\ell_2)^2-m_1^2+i \epsilon}\,,
\end{equation}
and the following soft scaling
\begin{equation}
\lim_{|\vec q|\to0}\frac{\delta((p_1-|\vec q|\tilde\ell_2-q)^2-m_1^2)}{(p_1+|\vec
  q|\tilde\ell_2)^2-m_1^2+i \epsilon}\sim \frac{\delta(2p_1\cdot
  \tilde\ell_2-q\cdot \hat\ell_2)}{|\vec q|^2 2p_1\cdot \tilde\ell_2}\sim
\mathcal O\left(\frac{1}{ |\vec{q}|^2}\right)\,,
\end{equation}
so that in the multi-soft limit $\ell_i=|q|\hat\ell_i$ and $|q|\to0$, each delta function adds a $\frac{1}{|\vec{q}|}$ factor in the soft expansion. Consequently, each term of the previous sum scales as
  \begin{equation}
  {\mathcal
  S}(\sigma,\gamma)_{\ell_{L\!+\!1}} (\delta( \dots ))^k\! \prod_{\sum_{j=1}^{k+1}  i_j=L} \!A_{i_j}^+  A^+_{L} \sim |\vec{q}|^{2L-2} \frac{1}{|\vec{q}|^{2k}} \frac{1}{|\vec{q}|^{L-1-k}}\frac{1}{|\vec{q}|^{L-1}}\sim \frac{1}{|\vec{q}|^{k}},
  \end{equation}
and the multiple soft behaviour for $|q|\to 0$ of the plus amplitude
is (for $k=0$)
\begin{align}
&\lim_{|q|\to0}  M^{\textrm{tree}(+)}_{L}(p_1,\ell_2, \dots
,\hat{\ell}_{L+1},-p_1') 
=(-1)^{L-1} \sum_{\sigma, \gamma\in\mathfrak S_{L-1}} {\mathcal
                S}(\sigma,\gamma)_{\ell_{L+1}}\cr
      &          \times\left. A^+_{L}(p_1,\sigma(2, \dots
                               ,L),\widehat{L+1},-p_1') \right |_{
                               |\vec{q}| \rightarrow 0} 
                               \left. A^+_{L}(p_1,\widehat{L+1},\gamma(2,
                               \dots ,L),-p_1')  \right |   _{
                               |\vec{q}| \rightarrow 0} \cr
                               &\sim |\vec{q}|^{0} \,.
\end{align}

We arrive to the conclusion that the leading soft contribution of $M^{\textrm{tree}(+)}$ is of order $|\vec{q}|^{0}$ and can be expressed as a product of leading soft Yang-Mills amplitudes. It means that $M^{\textrm{tree}(+)}$ will have the same universality properties as the soft Yang-Mills amplitudes. $M^{\textrm{tree}(-)}$ sharing the same property, we conclude that the universality property will be transmitted to all classical integrands.
%

%%%%%%%%%%%%%%%%%%%%%%%%%%%%%%%%%%%%%%%%%%%%%%%%%%%%%%%%%%%%%%
\section{Yang-Mills amplitudes and numerator factors}\label{sec:YMtrees}

Following the construction presented in~\cite{Bjerrum-Bohr:2020syg} we
can construct symmetric
numerators for scalar-gluon tree-level amplitudes.\footnote{An alternative
 construction of the numerators can be done using the tree-level BCJ
 master numerators derived from 10D pure-spinor
 formalism~\cite{Mafra:2015vca}.}
%-----------------------------------------------------------------------
\subsection*{The three-point amplitude and numerator factors}\label{sec:3pt}
For three-point scalar graviton amplitudes, we have,
\begin{equation} 
\begin{aligned}
N_1(p,\ell_2, -p')=&\,i\,\sqrt{2} \,\zeta_2\cdot p,\quad A_{1}(p,\ell_2,-p')=N_1(p,\ell_2, -p'),
\end{aligned}
\end{equation}
%-----------------------------------------------------------------------
\subsection*{The four-point amplitude and numerator factors}\label{sec:4pt}
For the four-point amplitude and numerator factors we have
\begin{equation} 
 \begin{aligned}\label{N4pt}\hskip-3.7cm
 N_2(p,\ell_2,\ell_3, -p')&= \frac i2 \Big(s_{2\,p} (\zeta_{2}\cdot\zeta_{3}) - 
  4(\zeta_2\cdot p) \zeta_3\cdot (p+ \ell_2)\Big)\,,
 \end{aligned}
\end{equation}
\begin{equation} 
 \begin{aligned}\label{A4pt}
{A}_{2}(p,\ell_2,\ell_3, -p')&=\frac{{N_2}(p,\ell_2,\ell_3, -p')}{\sss(2,p)}+\frac{{N_2}(p,\ell_2,\ell_3, -p')-{N_2}(p,\ell_3,\ell_2, -p')}{\sss(2,3)}\\ &=\frac{{N_2}(p,\ell_2,\ell_3, -p')}{\sss(2,p)}+\frac{{N_2}(p,[2,3],-p')}{\sss(2,3)}.
 \end{aligned}
\end{equation}
inspired by the compact notation of \cite{Agerskov:2019ryp}, and
where we have defined $s_{ip}\equiv(p+\ell_i)^2-m^2$, 
$s_{ij}\equiv(\ell_i+\ell_j)^2$, and 
\begin{multline}
  {N_2}(p,[2,3],-p')\equiv {N_2}(p,\ell_2,\ell_3, -p')-{N_2}(p,\ell_3,\ell_2, -p')=\cr\frac{1}{2} i ((s_{2p}-s_{3p}) \zeta_2\cdot\zeta_3-4 (\zeta_3\cdot\ell_2) (\zeta_2\cdot p)+4
   (\zeta_2\cdot\ell_3) (\zeta_3\cdot p))\,.
\end{multline}
%-----------------------------------------------------------------------
\subsection*{The five-point amplitude and numerator factors}\label{sec:5pt}
For the five-point amplitude and numerator factors we have
 \begin{multline}\label{N5pt}
N_3(p,\ell_2,\ell_3,\ell_4,-p')=\frac{-i}{\sqrt{2}}\Big(\frac{1}{3} \eee(3,4) (3 (\sss(2,3)+\sss(3,p))
  \eeek(2,p)-2 \sss(2,p)
  \eeek(2,{\ell_3}))\cr+\frac{1}{3} \sss(2,p)
  \eee(2,4) (2 \eeek(3,{\ell_2})+3
  \eeek(3,p))
+  \frac{1}{3} \sss(2,p) \eee(2,3)
  (2 \eeek(4,{\ell_2})+4 \eeek(4,{\ell_3})+3
  \eeek(4,p))\cr-2 \eeek(2,p)
  \eeek(3,({\ell_2}+p))
  \eeek(4,({\ell_2}+{\ell_3}+p))\Big)\,,
 \end{multline} 
with the colour-ordered amplitude
\begin{equation}
{\cal A}_{3}(p,\ell_2,\ell_3,\ell_4,-p')
 =
  \frac{{N_3}^{2,3,4}}{\sss(2,p)
  \ttt(2,3,p)}+\frac{{N_3}^{[2,3],4}}{\sss(2,3)
  \ttt(2,3,p)}+\frac{{N_3}^{2,[3,4]}}{\sss(3,4) \sss(2,p)}+\frac{{N_3}^{[2,[3,4]]}}{\sss(3,4)
  \ttt(2,3,4)}+\frac{{N_3}^{[[2,3],4]}}{\sss(2,3)
  \ttt(2,3,4)}\,, \end{equation}
where we have used the shorthand notation ${N_3}(p,\ell_2,\ell_3,\ell_4,-p')\equiv
N_3^{2,3,4}$ and ${N_3}^{[2,3],4}\equiv {N_3}^{2,3,4}-{N_3}^{3,2,4}$,
${N_3}^{2,[3,4]}\equiv {N_3}^{2,3,4}-{N_3}^{2,4,3}$,  
${N_3}^{[[2,3],4]}\equiv {N_3}^{[2,3],4}-{N_3}^{4,[2,3]}$ and
${N_3}^{[2,[3,4]]}\equiv {N_3}^{2,[3,4]}-{N_3}^{[3,4],2}$. We have
defined $s_{i,\dots ,j ,p}\equiv(p+\ell_i+\cdots+\ell_j)^2-m^2$,
$s_{i,\ldots ,j}\equiv(\ell_i+\cdots+\ell_j)^2$.

%-----------------------------------------------------------------------
\subsection*{The six-point amplitude and numerator factors}\label{sec:6pt}
We give the expression for the colour-ordered  six-point amplitude
(emission of four gluons from a massive scalar), The numerator factors are given on \href{https://nbviewer.org/github/pierrevanhove/Probe/blob/main/probe.ipynb}{this page}
 \begin{multline}\label{N5pt}
 {\cal A}_{4}(p,\ell_2,\ell_3,\ell_4,\ell_5,-p')=
   \frac{{N_4}^{2,3,4,5)}}{\sss(2,p) \ttt(2,3,p) \uuu(2,3,4,p)}
+\frac{{}{N_4}^{2,3,[4,5]}}{\sss(4,5) \sss(2,p)
  \ttt(2,3,p)}
+\frac{{}{N_4}^{2,[3,4],5)}}{\sss(3,4) \sss(2,p)
  \uuu(2,3,4,p)}
+\frac{{}{N_4}^{[2,3],4,5)}}{\sss(2,3) \ttt(2,3,p)
  \uuu(2,3,4,p)}\cr 
+\frac{{}{N_4}^{[2,3],[4,5]}}{\sss(2,3) \sss(4,5)
  \ttt(2,3,p)}
+\frac{{}{N_4}^{[[2,3],4],5)}}{\sss(2,3) \ttt(2,3,4)
  \uuu(2,3,4,p)}
+\frac{{}{N_4}^{[2,[3,4]],5)}}{\sss(3,4) \ttt(2,3,4)
  \uuu(2,3,4,p)}
+\frac{{N_4}^{2,[[3,4],5]}}{\sss(3,4) \ttt(3,4,5)
  \sss(2,p)}
+\frac{{N_4}^{2,[3,[4,5]]}}{\sss(4,5) \ttt(3,4,5)
  \sss(2,p)}\cr 
+\frac{{}{N_4}^{[[2,3],[4,5]]}}{\sss(2,3) \sss(4,5)
  \uuu(2,3,4,5)}
 +\frac{{}{N_4}^{[[[2,3],4],5]}}{\sss(2,3)
  \ttt(2,3,4)
  \uuu(2,3,4,5)}
  +\frac{{}{N_4}^{[[2,[3,4]],5]}}{\sss(3,4)
  \ttt(2,3,4)
  \uuu(2,3,4,5)}
  +\frac{{}{N_4}^{[2,[[3,4],5]]}}{\sss(3,4)
  \ttt(3,4,5)
  \uuu(2,3,4,5)}
  +\frac{{}{N_4}^{[2,[3,[4,5]]]}}{\sss(4,5)
  \ttt(3,4,5) \uuu(2,3,4,5)}
\,.
  \end{multline}
  %
%-----------------------------------------------------------------------
\subsection*{The seven-point amplitude and numerator factors}\label{sec:7pt}
We give the expression for the colour-ordered  seven-point amplitude
(emission of five gluons from a massive scalar), The numerator factors are given on \href{https://nbviewer.org/github/pierrevanhove/Probe/blob/main/probe.ipynb}{this page}
\begin{align}
&{\cal A}_{5}(p,\ell_2,\ell_3,\ell_4,\ell_5,\ell_6,-p')=\cr&
\frac{N_5^{2,3,4,5,6}}{\sss(2,p) \ttt(2,3,p) \uuu(2,3,4,p) \www(2,3,4,5,p)}
+\frac{N_5^{[2,3],4,5,6}}{\sss(2,3)   \ttt(2,3,p) \uuu(2,3,4,p)\www(2,3,4,5,p)}
+\frac{N_5^{2,[3,4],5,6}}{\sss(2,p) \sss(3,4) \uuu(2,3,4,p)\www(2,3,4,5,p)}
+\frac{N_5^{2,3,[4,5],6}}{\sss(2,p) \sss(4,5) \ttt(2,3,p) \www(2,3,4,5,p)}
\cr&
+\frac{N_5^{2,3,4,[5,6]}}{\sss(2,p) \sss(5,6) \ttt(2,3,p) \uuu(2,3,4,p)}
+\frac{N_5^{[2,3],4,[5,6]}}{\sss(2,3) \sss(5,6) \ttt(2,3,p)\uuu(2,3,4,p)}
+\frac{N_5^{2,[3,4],[5,6]}}{\sss(2,p) \sss(3,4) \sss(5,6)\uuu(2,3,4,p)}
+\frac{N_5^{2,3,[[4,5],6]}}{\sss(2,p) \sss(4,5)
     \ttt(2,3,p)\ttt(4,5,6)}
\cr&
+\frac{N_5^{2,3,[4,[5,6]]}}{\sss(2,p) \sss(5,6) \ttt(2,3,p)\ttt(4,5,6)}
+\frac{N_5^{[2,3],[4,5],6}}{\sss(2,3) \sss(4,5) \ttt(2,3,p)\www(2,3,4,5,p)}
+\frac{N_5^{[[2,3],4],5,6}}{\sss(2,3) \ttt(2,3,4) \uuu(2,3,4,p)\www(2,3,4,5,p)}
+\frac{N_5^{[2,[3,4]],5,6}}{\sss(3,4) \ttt(2,3,4) \uuu(2,3,4,p)\www(2,3,4,5,p)}
\cr&
+\frac{N_5^{2,[[3,4],5],6}}{\sss(2,p) \sss(3,4) \ttt(3,4,5)\www(2,3,4,5,p)}
+\frac{N_5^{2,[3,[4,5]],6}}{\sss(2,p) \sss(4,5) \ttt(3,4,5)\www(2,3,4,5,p)}
+\frac{N_5^{2,[[3,4],[5,6]}}{\sss(2,p) \sss(3,4) \sss(5,6)\uuu(3,4,5,6)}
+\frac{N_5^{[[2,3],[4,5]],6}}{\sss(2,3)\sss(4,5) \uuu(2,3,4,5)\www(2,3,4,5,p)}
          \end{align}
\begin{align*}
&+\frac{N_5^{[2,3],[6,[5,4]]}}{\sss(2,3) \sss(4,5) \ttt(2,3,p)\ttt(4,5,6)}
+\frac{N_5^{[2,3],[4,[5,6]]}}{\sss(2,3) \sss(5,6) \ttt(2,3,p)\ttt(4,5,6)}
+\frac{N_5^{[[2,3],4],[5,6]}}{\sss(2,3) \sss(5,6) \ttt(2,3,4)\uuu(2,3,4,p)}
+\frac{N_5^{[2,[3,4]],[5,6]}}{\sss(3,4) \sss(5,6) \ttt(2,3,4)\uuu(2,3,4,p)}
\cr&
+\frac{N_5^{[[[2,3],4],5],6}}{\sss(2,3) \ttt(2,3,4) \uuu(2,3,4,5)\www(2,3,4,5,p)}
+\frac{N_5^{[[2,[3,4]],5],6}}{\sss(3,4) \ttt(2,3,4) \uuu(2,3,4,5)\www(2,3,4,5,p)}
+\frac{N_5^{[2,[[3,4],5]],6}}{\sss(3,4) \ttt(3,4,5) \uuu(2,3,4,5)\www(2,3,4,5,p)}
+\frac{N_5^{[2,[3,[4,5]]],6}}{\sss(4,5) \ttt(3,4,5) \uuu(2,3,4,5)\www(2,3,4,5,p)}
\cr&
+\frac{N_5^{2,[[[3,4],5],6]}}{\sss(2,p) \sss(3,4) \ttt(3,4,5)\uuu(3,4,5,6)}
+\frac{N_5^{2,[[3,[4,5]],6]}}{\sss(2,p) \sss(4,5) \ttt(3,4,5)\uuu(3,4,5,6)}
+\frac{N_5^{2,[3,[[4,5],6]]}}{\sss(2,p) \sss(4,5) \ttt(4,5,6)\uuu(3,4,5,6)}
+\frac{N_5^{2,[3,[4,[5,6]]]}}{\sss(2,p) \sss(5,6) \ttt(4,5,6)\uuu(3,4,5,6)}
\cr&
+\frac{N_5^{[[2,3],[[4,5],6]]}}{\sss(2,3) \sss(4,5) \ttt(4,5,6)\www(2,3,4,5,6)}
+\frac{N_5^{[[2,3],[4,[5,6]]]}}{\sss(2,3) \sss(5,6) \ttt(4,5,6)\www(2,3,4,5,6)} 
+\frac{N_5^{[[[2,3],4],[5,6]]}}{\sss(2,3) \sss(5,6) \ttt(2,3,4)\www(2,3,4,5,6)}
+\frac{N_5^{[[2,[3,4]],[5,6]}}{\sss(3,4) \sss(5,6) \ttt(2,3,4)\www(2,3,4,5,6)}
\cr&
+\frac{N_5^{[[[2,3],[4,5]],6]}}{\sss(2,3) \sss(4,5) \uuu(2,3,4,5)\www(2,3,4,5,6)}
+\frac{N_5^{[2,[[3,4],[5,6]]]}}{\sss(3,4) \sss(5,6) \uuu(3,4,5,6)\www(2,3,4,5,6)}
+\frac{N_5^{[[[[2,3],4],5],6]}}{\sss(2,3) \ttt(2,3,4) \uuu(2,3,4,5)\www(2,3,4,5,6)}
+\frac{N_5^{[[[2,[3,4]],5],6]}}{\sss(3,4) \ttt(2,3,4) \uuu(2,3,4,5)\www(2,3,4,5,6)}
\cr&
+\frac{N_5^{[[2,[[3,4],5]],6]}}{\sss(3,4) \ttt(3,4,5) \uuu(2,3,4,5)\www(2,3,4,5,6)}
+\frac{N_5^{[[2,[3,[4,5]]],6]}}{\sss(4,5) \ttt(3,4,5) \uuu(2,3,4,5)\www(2,3,4,5,6)}
+\frac{N_5^{[2,[[[3,4],5],6]]}}{\sss(3,4) \ttt(3,4,5) \uuu(3,4,5,6)\www(2,3,4,5,6)}
+\frac{N_5^{[2,[[3,[4,5]],6]]}}{\sss(4,5) \ttt(3,4,5) \uuu(3,4,5,6)\www(2,3,4,5,6)}
\cr&
+\frac{N_5^{[2,[3,[[4,5],6]]]}}{\sss(4,5) \ttt(4,5,6) \uuu(3,4,5,6)\www(2,3,4,5,6)}
+\frac{N_5^{[2,[3,[4,[5,6]]]]}}{\sss(5,6) \ttt(4,5,6) \uuu(3,4,5,6)\www(2,3,4,5,6)}\,.
\end{align*}

\end{document}